\documentclass[11pt]{article}
\usepackage{jheppub}
\usepackage{graphicx,color}
\usepackage[dvipsnames]{xcolor}
\usepackage{braket}
\usepackage{soul}
\usepackage{amssymb,amsmath}
\usepackage{bbm}
\usepackage{array}
\usepackage{simpler-wick}
\usepackage{verbatim}
\usepackage{anyfontsize}
\usepackage{dsfont}
\usepackage{tikz}
\usepackage{subcaption}
\usepackage{sidecap}
\usepackage{ytableau}
\sidecaptionvpos{figure}{t}
\usepackage[english]{babel}
\usepackage{enumitem}  
\usepackage[export]{adjustbox}
\usepackage{bbold}
\allowdisplaybreaks
\usepackage[]{hyperref}
\hypersetup{
    colorlinks = true,
    allcolors = {Periwinkle}
}
\allowdisplaybreaks

\newcommand{\nn}{\nonumber\\}

\def\ib{{\bar{\imath}}}
\def\jb{{\bar{\jmath}}}
\def\id{{\mathbbm{1}}}

\def\d{\mathrm{d}}
\def\i{\mathrm{i}}
\def\e{\mathrm{e}}
\def\cc{\mathrm{c.c.}}

\title{Open-closed 3d gravity as a random ensemble}

\author{Daniel L. Jafferis,}
\author{Liza Rozenberg,}
\author{Diandian Wang}

\affiliation{Jefferson Physical Laboratory, Harvard University, Cambridge, MA 02138, USA}
\emailAdd{jafferis@g.harvard.edu}
\emailAdd{erozenberg@g.harvard.edu}
\emailAdd{diandianwang@fas.harvard.edu}

\date{\today}

\begin{document}

\abstract{
    We investigate an ensemble of boundary CFTs within the framework of a tensor model recently constructed to model 3d quantum gravity. The incorporation of CFT borders introduces new elements to the gravity theory. In particular, it leads to an open-closed extension of Virasoro TQFT, which in the classical limit gives rise to 3d gravity with tensionful end-of-the-world branes. It also provides predictions for off-shell manifolds with bordered asymptotic boundaries, such as the annulus wormhole. As an application, we construct a purely open variant of the tensor model to study a purely open bootstrap problem in the context of CFT triangulation. We also briefly discuss the extension to non-orientable CFTs.  
}

\maketitle

\section{Introduction}

Three-dimensional gravity (3d gravity) has long served as a fertile ground for exploring quantum gravity and holography. Compared to gravity in higher dimensions, it enjoys remarkable simplicity due to the absence of local degrees of freedom, rendering it classically equivalent to a Chern-Simons theory \cite{Achucarro:1986uwr,Witten:1988hc}. Nevertheless, the triviality in the local geometry is compensated by the richness of 3d topology. Upon quantization, one obtains a topological quantum field theory (TQFT) called Virasoro TQFT \cite{Collier:2023fwi,Collier:2024mgv}, which constitutes an irrational counterpart to the Chern-Simons/Wess-Zumino-Witten correspondence \cite{Witten:1988hf}, with an inner product given by Verlinde \cite{Verlinde:1989ua}. The gravitational path integral of 3d gravity has provided intriguing insights about the spectrum of a putative dual description \cite{Maloney:2007ud,Keller:2014xba}, which has been shown to exhibit features of random matrix theory \cite{Cotler:2020ugk,Cotler:2020hgz,DiUbaldo:2023qli,Boruch:2025ilr,deBoer:2025rct}. Moreover, wormholes in 3d gravity capture statistical information about ensembles of CFTs \cite{Belin:2020hea,Anous:2021caj,Chandra:2022bqq,deBoer:2023vsm,deBoer:2024mqg,Wang:2025bcx} that are related to universal features obtainable from CFT bootstrap \cite{Kraus:2016nwo,Cardy:2017qhl,Collier:2019weq,Belin:2021ryy,Kusuki:2021gpt,Numasawa:2022cni}.

It has recently been understood that the full partition function of pure AdS$_3$ gravity, including the sum over all 3-manifold topologies, can be obtained as the topological expansion of a tensor and matrix integral \cite{Belin:2023efa,Jafferis:2024jkb}. The integration variables can be interpreted as CFT$_2$ data, namely conformal weights and OPE coefficients, and the integration measure imposes the conformal bootstrap constraints. This duality also provides a three-dimensional version of the connection between matrix models and two-dimensional quantum gravity \cite{Saad:2019lba}, along the lines of \cite{Turaev:1992hq,Freidel:2005qe,Gurau:2011xp,Rivasseau2016}.

In this paper, we expand the correspondence to a gravitational theory that admits dynamical boundaries referred to as end-of-the-world (EOW) branes. On the tensor model side, this introduces new degrees of freedom associated to the CFT data of a conformal boundary condition, together with additional terms in the tensor potential related to the richer collection of crossing relations for boundary conformal field theories (BCFTs). This is somewhat analogous to the addition of $N_f$ vectors to matrix models, corresponding to boundaries of the dual 2d worldsheets. In the 3d case, it remains an interesting open question whether there are analogs of the large-$N_f$ limit in which the theory simplifies, given the greater complexity of BCFT data.

The new degrees of freedom together with the extended tensor potential manifest themselves in the bulk as an open-closed generalization of Virasoro TQFT \cite{Eberhardt:2023mrq}. Unlike the purely closed case, the gluing rules of a 3d open-closed TQFT involve gluing along surfaces that are themselves bordered 2d manifolds \cite{Lazaroiu:2000rk,Lauda:2005wn,Moore:2006dw}. We formulate the open-closed Virasoro TQFT based on the Moore-Seiberg consistency conditions for BCFT, and show how it enriches the 3d sum over manifolds in the topological expansion of the new tensor model. 

In addition to new tensors, there is also a new matrix introduced to capture the statistics of the boundary operator spectrum (or the ``open-string" spectrum) of the BCFT ensemble. These lead to predictions for partition functions of off-shell 3d manifolds, such as the annulus wormhole (annulus times an interval), the open analog of the torus wormhole \cite{Cotler:2020ugk}.

The incorporation of additional bootstrap constraints also highlights an interesting application of ensemble models whose measure imposes constraints: Consider two collections of data and constraints. In the simplest situations, the data would be the same, and only the constraints would differ. Suppose one wants to determine whether the resulting bootstrap problems have the same set of solutions, particularly in situations in which it may not be obvious that one set of constraint equations implies the other. One can then define associated ensemble integrals over the data with a potential given by the squares of the constraints, and check if the associated topological expansions of the models agree. The new tensor model therefore provides a way to investigate the interplay between various bootstrap constraints. In fact, using 3d TQFT consistency as the guide, we clarify the role of one of the crossing relations in the literature and demonstrate its redundancy (in Section~\ref{ssec:ocTQFT}).

An application of the BCFT tensor model is to a purely open bootstrap, i.e., one that involves only the boundary spectrum and OPE coefficients, in the spirit of \cite{Hung:2019bnq,Brehm:2021wev,Chen:2022wvy,Cheng:2023kxh,Chen:2024unp,Brehm:2024zun}. Instead of defining a 2d CFT via the usual operator data by computing partition functions via a decomposition into pairs of pants, one can use purely boundary data to evaluate a triangulated partition function as follows. Given a triangulation of a Riemann surface, excise the vertices and impose a given fixed boundary condition on all of the resulting holes. The limit in which the holes shrink to zero size reproduces the original partition function up to a factor of $(\e^{\frac{\pi c}{6 \epsilon}}g)^V$, where $V$ is the number of vertices of the triangulation, $g$ is the $g$-function corresponding to the boundary condition, and $\epsilon$ is the size of each hole. Moreover, the triangulated partition function can be evaluated in terms of the boundary three-point functions and weights. 

Given a local CFT$_2$ and any local conformal boundary condition, the above construction applies. It is an interesting and highly non-trivial question whether all solutions of the purely boundary bootstrap, with data consisting of only the boundary weights and structure constants, subject only to boundary four-point crossing (the only bootstrap constraint in the purely open case), come from a boundary condition of a local CFT. One piece of heuristic evidence is that boundary four-point crossing is sufficient to imply that all triangulations give the same result in the shrinking limit, and very dense triangulations seem to imply a form of 2d locality. The bulk CFT$_2$ data, if it exists, can be reconstructed by computing triangulated torus and higher-genus partition functions. 

The question for the associated purely boundary data tensor model is whether the topological expansion for CFT partition functions agrees with that of the ``closed'' CFT tensor model, i.e., with pure 3d gravity without contributions of EOW branes. The tensor and matrix diagrams that contribute to the purely open model are a subset of those of a full BCFT model that we analyze in detail in this paper. In particular, our results show that every diagram evaluates to the 3d gravity partition function on an associated 3-manifold with EOW branes. Therefore, the above question reduces to whether the contributions from manifolds with extra interior branes or higher genus branes filling in the holes of the triangulation are suppressed. 

To answer this question, it is useful to analyze the contributions from 3d manifolds with different topologies. In particular, there are singular contributions associated to non-hyperbolic manifolds, even in the closed case, i.e., without EOW branes: For example, adding a handle (sphere times an interval) to a hyperbolic manifold can lead to a divergence owing to the limit where the handle becomes very long. The analog of this example in the open case is ``half'' of the handle: Take a $\mathbb{Z}_2$ quotient of the handle and place a spherical EOW brane at the fixed points. Then glue the only remaining boundary to a hyperbolic manifold (with a 3-ball removed). The divergence again comes from the limit where this half handle becomes long, but the situation is potentially more manageable. One might expect that adding a local counter term whenever a spherical EOW brane appears could regulate the partition functions on such manifolds. While we leave a detailed investigation of this possibility to future work, it serves as both a motivation for and an application of the BCFT tensor model.

As another natural generalization, we also discuss the incorporation of non-orientable 2d CFTs and correspondingly the additional bootstrap conditions. These are CFTs that have a reflection symmetry and admit a definition on the crosscap. This further enriches the topological sum by including non-orientable 3d manifolds. In particular, we explicitly obtain some new TQFT building blocks and discuss their implications for the ensemble interpretation.

This paper is organized as follows. In Section~\ref{sec:cstrs}, we review the BCFT data and bootstrap. In Section~\ref{sec:model}, we construct the tensor-matrix model incorporating the new degrees of freedom and bootstrap constraints. Section~\ref{sec:3d} presents the 3d interpretation of each element of the new model, including the data, propagators, and interaction vertices. In particular, we explain how this bulk interpretation gives rise to an open-closed Virasoro TQFT. Section~\ref{sec:pureopen} presents the aforementioned application to CFT triangulation by restricting the model to a purely open version. In Section~\ref{sec:xcft}, we discuss how to incorporate non-orientability. 

\paragraph{Note added.} While finalizing the paper, we became aware of \cite{Hung:2025vgs}, which contains some overlap with our work.

\section{Review of BCFT}\label{sec:cstrs}

\subsection{Data}\label{ssec:bdata}
In BCFT, besides conformal weights $( h_i, \bar{h}_i)$ and OPE coefficients $C_{ijk}$, additional data are necessary to fully characterize the theory. These include boundary conditions, boundary operator spectra, boundary OPE coefficients, and structure constants that govern the interactions between bulk and boundary operators. 

Following \cite{Wang:2025bcx}, we will use the notation $\phi_i$ for bulk operators which are  labeled with lower-case letters ($i$, $j$, ...) and $\phi_{I}^{(ab)}$ for boundary operators which are labeled with upper-case letters ($I$, $J$, ...). The superscripts $(ab)$ keep track of the boundary conditions on the two sides of the boundary operator insertion, which can generally be different. With this notational remark in mind, all the data needed to specify a BCFT are:
\begin{itemize}
    \item $g_a$: the $g$-function corresponding to the boundary condition $a$, which is defined to be the partition function of a disk with boundary condition labeled by $a$ imposed at the circular border;
    \item $(h_i,\bar{h}_i)$: conformal weights of bulk operators;
    \item $h_I^{(ab)}$: conformal weights of boundary operators in the Hilbert space $\mathcal{H}_{\rm closed}$;
    \item $C_{ijk}$: bulk-to-bulk OPE coefficients, or the three-point function of $\phi_i$, $\phi_j$, and $\phi_k$ on the sphere in the Hilbert space $\mathcal{H}^{(ab)}_{\rm open}$ (or $\mathcal{H}^{ab}$ for short);
    \item $B^{(abc)}_{IJK}$: boundary-to-boundary OPE coefficients, or the three-point function of $\phi_I^{(ab)}$, $\phi_J^{(bc)}$, and $\phi_K^{(ca)}$ on the disk, whose circular border is partitioned into three intervals by these operator insertions, with boundary conditions $a$, $b$, and $c$, respectively;
    \item $D^{(a)}_{iI}$: bulk-to-boundary OPE coefficients, or the two-point function of $\phi_i$ and $\phi_{I}^{(aa)}$ on the disk.
\end{itemize}
Pictorially, we represent the three types of OPE coefficients as
\begin{align}\label{eq:datapics}
   C_{ijk} \longrightarrow \vcenter{\hbox{\includegraphics[height=1.5cm]{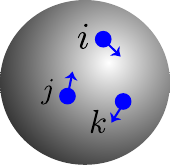}}}~,
   \qquad B_{IJK}^{(abc)} \longrightarrow \vcenter{\hbox{\includegraphics[height=1.5cm]{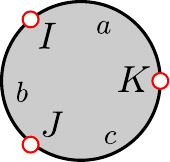}}}~,
   \qquad D_{iI}^{(a)} \longrightarrow \vcenter{\hbox{\includegraphics[height=1.5cm]{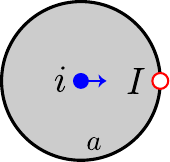}}}~.
\end{align}
At this point, the little arrows can be thought of as a reminder of the fact that the bulk operators have spin, but they will be given a more precise interpretation later. For bulk operators, we choose the standard normalization of the two-point function on the sphere such that
\begin{align}
    \langle\phi_i(z_i,\bar{z}_i) \phi_j(z_j,\bar{z}_j)\rangle=\frac{\delta_{ij}}{(z_i-z_j)^{2h_i}(\bar{z}_i-\bar{z}_j)^{2\bar{h}_i}},
\end{align}
where $z_i$ and $\bar{z}_i$ are coordinates of the complex plane. With this normalization, $\delta_{ij}$ is used to raise and lower $i$ indices, so there is no difference between upper and lower $i$ indices. For boundary operators, we choose the canonical normalization of the two-point function of boundary operators such that 
\begin{equation}\label{eq:bdy2ptnorm}
\langle\phi_I^{(a b)}(x_I) \phi_J^{(b a)}(x_J)\rangle\equiv \frac{\mathfrak{g}_{IJ}^{(a b)}}{|x_I-x_J|^{2 h_I}}= \frac{\sqrt{g_a g_b} \delta_{IJ}}{|x_I-x_J|^{2 h_I}},
\end{equation}
where $x$ is the coordinate of the real line, and   $\mathfrak{g}_{IJ}^{(a b)}$ acts as the metric used to raise and lower the $I$ indices. For example,
\begin{equation}
B_{IJK}^{(a b c)}\equiv\sum_{K^{\prime}} B_{IJ}^{(a b c) K^{\prime}} \mathfrak{g}_{KK^{\prime}}^{(a c)}
=
B_{IJ}^{(a b c) K} \sqrt{g_a g_c}.
\end{equation}
Conformal symmetry implies
\begin{align}\label{eq:sym_conf}
    C_{\sigma(i)\sigma(j)\sigma(k)}&=(-1)^{(s_i+s_j+s_k)(1-\operatorname{sgn}(\sigma))/2}C_{ijk},\quad     B_{IJK}^{(abc)}=B_{JKI}^{(bca)}=B_{KIJ}^{(cab)},
\end{align}
where $\sigma$ is a permutation and $\operatorname{sgn}(\sigma)$ is positive (negative) one for even (odd) permutations. Notice that, unlike $C$, swapping two indices of $B$ gives an independent quantity (though they are related by complex conjugation, as we explain momentarily).  Reflection positivity imposes
\begin{align}\label{eq:RP}
    C^*_{ijk}&=(-1)^{s_i+s_j+s_k}C_{ijk},\quad B_{IJK}^{*(abc)}=B_{KJI}^{(acb)},\quad 
    D^{*(a)}_{iI}=D^{(a)}_{iI}.
\end{align}
To see this, consider for example the four-point function
\begin{align}
    \vcenter{\hbox{\includegraphics[height=2cm]{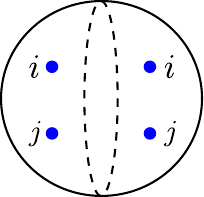}}}=\sum_k C_{ijk}C_{jik}\,\mathcal{G}_k, 
\end{align}
where, for every $k$, $\mathcal{G}_k$ is a positive function that depends on the positions of the insertions. Requiring this to be non-negative for all configurations (cross ratios) implies non-negativity of each summand, i.e., $C_{ijk}C_{jik} \ge 0$. We then conclude that $C_{ijk}$ and $C_{jik}$ must have opposite phases. Since we also know $|C_{ijk}|=|C_{jik}|$ from \eqref{eq:sym_conf}, we conclude that\footnote{In higher dimensions, it is common to focus on parity-even tensor structures of the three-point function, in which case $\mathcal{G}_k$ is positive for even total spin and negative for odd total spin, making $C_{ijk}$ always real \cite{Kravchuk:2021kwe}.}
\begin{align}
    C_{ijk} = C_{jik}^*.
\end{align}
The derivation is slightly more involved for $B$. From
\begin{align}
    \vcenter{\hbox{\includegraphics[height=2cm]{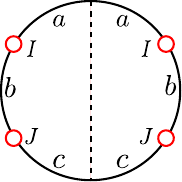}}}\ge 0 \implies B_{IJK}^{(abc)}B_{JIK}^{(cba)}\ge0,
\end{align}
we infer that $B_{IJK}^{(abc)}$ and $B_{JIK}^{(cba)}$ have opposite phases. To arrive at $B_{IJK}^{(abc)}=B_{JIK}^{*(cba)}$, we should consider the more general setting where the left half-disk is a linear combination of the state shown and the state with $I$ and $J$ swapped; requiring the corresponding disk correlator to be non-negative for arbitrary cross ratios and linear combinations yields the desired relation.

Finally, for $D$, consider
\begin{align}
    \vcenter{\hbox{\includegraphics[height=1.8cm]{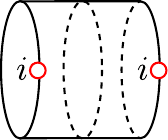}}}\ge 0 \implies D_{iI}^{(a)}D_{iI}^{(a)}\ge0,
\end{align}
so $D_{iI}^{(a)}$ must be real.

\subsection{Bootstrap}
For CFTs defined on closed orientable Riemann surfaces, there are two independent bootstrap constraints, namely crossing symmetry of four-point functions on the sphere and modular covariance of one-point functions on the torus \cite{Moore:1988uz,Moore:1988qv}. In the case of BCFTs, these constraints are extended by four additional conditions \cite{Cardy:1991tv,Lewellen:1991tb}. In what follows, we provide a brief review of these six constraints. We will express them in terms of crossing kernels rather than conformal blocks. This is because the kernels are often more convenient to work with and in our case explicitly known \cite{Ponsot:1999uf,Ponsot:2000mt,Teschner:2012em,Teschner:2013tqy}, so we can avoid dealing with unknown functions.

\subsubsection*{Four bulk operators on the sphere}

The first constraint comes from the crossing symmetry of four bulk operators on the sphere. Pictorially, it states
\begin{align}\label{eq:cstr1block}
    \sum_m C_{ijm} C_{mkl} \Bigg| \vcenter{\hbox{\includegraphics[scale=0.5]{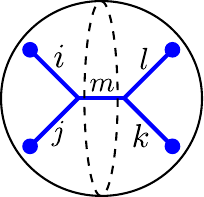}}} \Bigg\rangle = \sum_m C_{jkm} C_{mli} \Bigg|  \vcenter{\hbox{\includegraphics[scale=0.5]{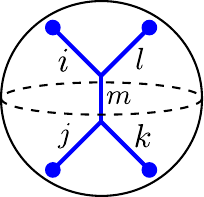}}} \Bigg\rangle.
\end{align}
The two skeleton diagrams that appear in this equation represent two different conformal block decompositions of the four-point function, namely the $s$-channel and $t$-channel decompositions. 
The dashed circle in each diagram represents a state cut, where a resolution of the identity is inserted, and the solid blue line going through the circle represents these internal states. 

To work with crossing kernels rather than the conformal blocks, we rewrite this constraint in the following way. First, write the $t$-block as a linear combination of $s$-blocks, using the spherical crossing kernel, defined via
\begin{equation}\label{eq:Fdef}
    \mathcal{F}_t(P_t | 1-z)=\int_0^\infty {\d P_s} \,\mathbb{F}_{P_t P_s}
    \begin{bmatrix}
    P_k & P_l \\
    P_j & P_i
    \end{bmatrix}
    \mathcal{F}_s(P_s | z),
\end{equation}
where $\mathcal{F}_s$ and $\mathcal{F}_t$ are the $s$-channel and $t$-channel conformal blocks, respectively, $z$ is the cross ratio, and $\mathbb{F}$ is the Ponsot-Teschner fusion kernel \cite{Ponsot:1999uf,Ponsot:2000mt}. Projecting the equation onto an $s$-channel block labeled by $P$ then leads to
\begin{align}\label{eq:M1}
    \mathcal{M}_{1,ijkl}(P,\bar{P})\equiv \sum_m \left(C_{ijm} C_{mkl} \delta^{2}(P-P_m) - C_{jkm} C_{mli} \left|\mathbb{F}_{mP}
    \!\begin{bmatrix}
        k & l \\
        j & i
    \end{bmatrix}
    \right|^2\right) = 0,
\end{align}
where we have used a simplified notation
\begin{align}
    \mathbb{F}_{P_mP}\!\begin{bmatrix}
        P_k & P_l \\
        P_j & P_i
    \end{bmatrix}
    \longrightarrow \mathbb{F}_{mP}\!\begin{bmatrix}
        k & l\\
        j & i
    \end{bmatrix}.
\end{align}
Similar notational simplifications will be employed throughout the paper.

\subsubsection*{One bulk operator on the torus}
The second constraint comes from modular covariance of one bulk operator on the torus. It is given by the equation
\begin{align}\label{eq:cstr2block}
    \sum_j C_{ijj}  \Bigg| \vcenter{\hbox{\includegraphics[scale=0.6]{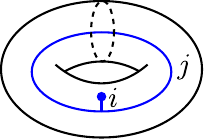}}} \Bigg\rangle = \sum_j C_{ijj} \Bigg|  \vcenter{\hbox{\includegraphics[scale=0.6]{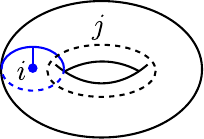}}} \Bigg\rangle.
\end{align}
As before, the blue lines form skeleton diagrams, now lying on the torus, and the black dashed circles are the places where a complete basis of states labeled by $j$ are inserted. Writing the second block as a linear combination of the first and projecting it onto a single block labeled by $P$ again gives an equation involving only the crossing kernel, absent the blocks themselves:
\begin{align}\label{eq:cstr2}
\mathcal{M}_{2,i}(P,\bar{P})\equiv\sum_j C_{ijj} \delta^{2}(P-P_j)-C_{ijj}\left|\mathbb{S}_{P j}[i]\right|^2=0.
\end{align}
Here $\mathbb{S}_{P j}[i]$ is a shorthand for $\mathbb{S}_{P P_j}[P_i]$, the modular crossing kernel.

\subsubsection*{Four boundary operators on the disk}
After reviewing the two constraints for CFTs without borders, let us now consider the first new constraint needed for BCFTs. This comes from two ways of computing the function of four boundary operators on the disk. It is the open-sector analog of the spherical crossing symmetry. In terms of conformal blocks, it reads
\begin{align}\label{eq:cstr3block}
    \sum_{M\in\mathcal{H}^{ac}} \frac{1}{\sqrt{g_ag_c}}B^{(abc)}_{IJM}B^{(cda)}_{KLM} \Bigg| \vcenter{\hbox{\includegraphics[scale=0.5]{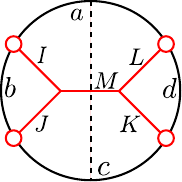}}} \Bigg\rangle = \sum_{M\in\mathcal{H}^{bd}} \frac{1}{\sqrt{g_bg_d}}B^{(dab)}_{LIM}B^{(bcd)}_{JKM} \Bigg|  \vcenter{\hbox{\includegraphics[scale=0.5]{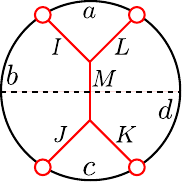}}} \Bigg\rangle.
\end{align}
Notice that this is a slightly more intricate situation than crossing on the sphere. The border of the disk is partitioned into four intervals by the operator insertions, each generally having a different boundary condition. The dashed line is an open state cut, where a complete set of states labeled by $M$ is inserted ($\phi_M^{(ac)}\in\mathcal{H}^{ac}$ on the left and $\phi_M^{(bd)}\in\mathcal{H}^{bd}$ on the right). The presence of the factors of $g_s$ is a result of a choice of the normalization for the two-point function of boundary operators on the disk \eqref{eq:bdy2ptnorm}.

Similar to crossing on the sphere, the $t$-channel here can be written as a linear combination of $s$-channel blocks. The kernel in fact is the same as one holomorphic factor of the spherical crossing kernel. Projecting onto an $s$-channel block with the internal weight labeled by $P$, the constraint can be written as
\begin{align}
\mathcal{M}^{(abcd)}_{3,IJKL}(P)\equiv&\frac{1}{\sqrt{g_ag_c}}\sum_{M \in \mathcal{H}^{ac}} B_{IJM}^{(a b c) } B_{KLM}^{(c d a)}\delta\left(P-P_M\right)\nn
-&
\frac{1}{\sqrt{g_dg_b}}\sum_{M \in \mathcal{H}^{bd}} B_{LIM}^{(d a b)} B_{JK M}^{(b c d)}
\mathbb{F}_{MP}
\begin{bmatrix}
J & I \\
K & L
\end{bmatrix}
=0.
\end{align}

\subsubsection*{One bulk and two boundary operators on the disk}
The fourth constraint comes from considering one bulk and two boundary operators on the disk. It requires
\begin{align}\label{eq:cstr4block}
    \sum_{M\in\mathcal{H}^{aa}} \frac{1}{g_a}D^{(a)}_{iM}B^{(aab)}_{MIJ} \Bigg| \vcenter{\hbox{\includegraphics[scale=0.5]{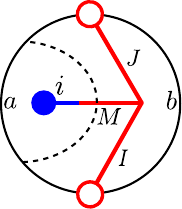}}} \Bigg\rangle = \sum_{M\in\mathcal{H}^{bb}} \frac{1}{g_b}D^{(b)}_{iM}B^{(bba)}_{MIJ} \Bigg|  \vcenter{\hbox{\includegraphics[scale=0.5]{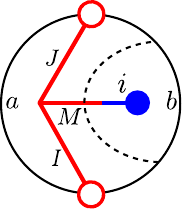}}} \Bigg\rangle.
\end{align}
The two blocks are similar, but due to the fact that boundary conditions $a$ and $b$ are generally different, taking the bulk operator $i$ to a boundary operator on the interval with boundary condition $a$ using the bulk-to-boundary OPE is different from taking it in the other direction to the interval labeled by $b$. Projecting the equation onto a block with internal weight $P$, the constraint takes the form
\begin{equation}
    \mathcal{M}^{(ab)}_{4,iIJ}(P)=0,
\end{equation}
where
\begin{align}
    \mathcal{M}^{(ab)}_{4,iIJ}(P)&\equiv    \frac{1}{g_a}\sum_{M\in\mathcal{H}^{aa}}D^{(a)}_{iM}B^{(aab)}_{MIJ}\delta(P-P_M) 
    \\
    &-
    \frac{1}{g_b}\sum_{M\in\mathcal{H}^{bb}}D^{(b)}_{iM}B^{(bba)}_{MJI}
    \int_0^\infty \d P' \d P'' 
    \mathbb{F}_{M P'}
    \begin{bmatrix}
        \ib & I \\
        i & J
    \end{bmatrix}
    \mathbb{F}_{P' P''}
    \begin{bmatrix}
        \ib & J \\
        I & i
    \end{bmatrix}
    \mathbb{F}_{P'' P}
    \begin{bmatrix}
        i & \ib \\
        I & J
    \end{bmatrix}
    \mathbb{B}_{\ib I}^{P'}
    \mathbb{B}^{Ii}_{P''}.\nonumber
\end{align}
In all previous constraints, only the fusion matrix $\mathbb{F}$ and modular matrix $\mathbb{S}$ have appeared. Here we see a new object $\mathbb{B}$:
\begin{equation}
    \mathbb{B}{}^{12}_{3}=\mathrm{e}^{\i\pi\left(-h_1-h_2+h_3\right)}, 
    \quad \mathbb{B}_{12}^{3}\equiv
    (\mathbb{B}{}^{12}_{3})^{*}.
\end{equation}
This corresponds to an operation known as braiding. Together, spherical crossing, modular crossing and braiding constitute all the basic moves needed to transform between any conformal blocks.

\subsubsection*{Two bulk and one boundary operators on the disk}
The fifth constraint comes from the consideration of two bulk operators and one boundary operator on the disk. Equivalence of the two conformal block decompositions states that
\begin{align}\label{eq:cstr5block}
    \sum_{m} C_{ijm} D_{mI}^{(a)} \Bigg| \vcenter{\hbox{\includegraphics[scale=0.5]{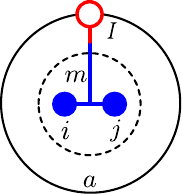}}} \Bigg\rangle = \frac{1}{g_a^2}  \sum_{M,N\in\mathcal{H}^{aa}} B_{MNI}^{(aaa)} D_{iM}^{(a)} D_{jN}^{(a)} \Bigg|  \vcenter{\hbox{\includegraphics[scale=0.5]{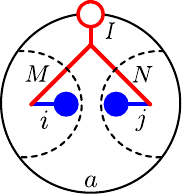}}} \Bigg\rangle.
\end{align}
In the first conformal block decomposition, the two bulk operators are taken to approach each other first, fusing into another bulk operator via the bulk-to-bulk OPE. The resulting bulk operator is then taken to the boundary via the bulk-to-boundary OPE. In the second conformal block decomposition, each bulk operator is taken to the boundary first via the bulk-to-boundary OPE, and the resulting boundary operators are then evaluated using the boundary-to-boundary OPE. Projecting onto the first channel with the internal weight labeled by $P$, we can rewrite this constraint as $M^{(a)}_{5,ijI}(P,\bar{P})=0$ with
\begin{align}
   &M^{(a)}_{5,ijI}(P,\bar{P})\\
   \equiv\,&
   \sum_{m} C_{ijm} D_{mI}^{(a)} \delta^2(P-P_m)  -   \sum_{M,N} \frac{1}{g_a^2}B_{MNI}^{(aaa)} D_{iM}^{(a)} D_{jN}^{(a)} \mathbb{K}_5[P_M,P_N;P,\bar{P}; P_I,P_i,P_j] ,\nonumber
\end{align}
where the crossing kernel is given by
\begin{align}
    \mathbb{K}_5[P_I,P_J;P_k,\bar{P}_k; P_K,P_i,P_j] &\equiv \int_0^{\infty} \d P \,\e^{\i\frac{\pi}{2} (h_K + h_I -h_J - h_i +\bar{h}_i+h_j+\bar{h}_j+h_k-\bar{h}_k-2h)}\nn
    & \times
    \mathbb{F}_{J P} \begin{bmatrix}
        K & \jb \\ I & j 
    \end{bmatrix} 
    \mathbb{F}_{I k} \begin{bmatrix}
        \ib & P \\ i & j
    \end{bmatrix} 
    \mathbb{F}_{P \bar{k}} \begin{bmatrix}
        \ib & \jb \\ k & K
    \end{bmatrix}.
\end{align}

\subsubsection*{Two boundary operators on the annulus}
Finally, the sixth constraint concerns two boundary operators on the annulus. In \cite{Lewellen:1991tb} where the full set of constraints was first presented, it was remarked that this constraint may be redundant. Some evidence was given in \cite{Fjelstad:2006aw}. Nevertheless, from the 2d perspective, it is not obvious how it follows from the other constraints. In this section, we will include this constraint nonetheless, but we will be able to understand this redundancy from a 3d perspective in a later section. 

The constraint takes the form
\begin{align}\label{eq:cstr6block}
    \sum_{m} D_{mI}^{(a)} D_{mJ}^{(b)} \Bigg| \vcenter{\hbox{\includegraphics[scale=0.6]{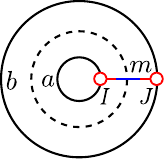}}} \Bigg\rangle = \sum_{K,L\in\mathcal{H}^{ab}}\frac{1}{g_ag_b} B_{IKL}^{(aab)} B_{JKL}^{(bba)}\Bigg|  \vcenter{\hbox{\includegraphics[scale=0.6]{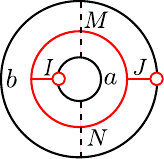}}} \Bigg\rangle.
\end{align}
In the picture on the left, a bulk state is inserted at the circular cut, while in the picture on the right, boundary states are inserted at the interval cuts. We therefore refer to the left and right block decompositions as closed and open channels, respectively.

Projecting the equation onto a bulk channel with the internal weight labeled by $P$, the constraint becomes $M^{(ab)}_{6,IJ}(P,\bar{P})=0$ with
\begin{align}\label{eq:ctsr6}
    &M^{(ab)}_{6,IJ}(P,\bar{P}) \\
    \equiv\,&\sum_{m} D_{mI}^{(a)} D_{mJ}^{(b)} \delta^2(P-P_m)  - \sum_{K,L\in\mathcal{H}^{ab}}\frac{1}{g_ag_b} B_{IKL}^{(aab)} B_{JKL}^{(bba)}\mathbb{K}_{6} [P_M,P_N;P, \bar{P};P_I, P_J] , \nonumber
\end{align}
where the crossing kernel $\mathbb{K}_{6}$ is unspecified. It turns out that we do not need to know it explicitly as a result of this constraint being redundant. This will be explained in detail in Section~\ref{sec:3d}.

\section{Constructing the tensor-matrix model}\label{sec:model}

Following \cite{Belin:2023efa, Jafferis:2024jkb}, to construct the tensor-matrix model, we square each of the six constraints and add them to form the tensor model potential. This would give a potential that is minimized by BCFT data satisfying all the bootstrap conditions. The first two constraints are concerned with only the closed sector, in the sense that only $C$'s and $\phi_i$'s are involved; the third involves only the open sector since only $B$'s and $\phi_I^{(ab)}$'s are involved; the remaining ones involve both the open sector and the closed sector. In the following, we first review the potential coming from the closed sector \cite{Belin:2023efa,Jafferis:2024jkb} before introducing each new one.

The open-closed model we now construct is defined by
\begin{equation}\label{eq:Zopen-closed}
\mathcal{Z}=\prod_{s \in \mathbb{Z}} \int 
    \mathcal{D} [\Delta_s,h_I,C_{i j k},B^{(abc)}_{IJK},D^{(a)}_{iI} ,A^{(a)}_{i}] \,
    \e^{-V_0(\Delta_s,h_I)-\frac{1}{\hbar} V(\Delta_s, h_I,C_{i j k},B^{(abc)}_{IJK},D^{(a)}_{iI} ,A^{(a)}_{i})},
\end{equation}
where all indices run over only non-vacuum states. Here $A_{i}^{(a)}$ is defined to be $D^{(a)}_{i\id}$. 

As in \cite{Jafferis:2024jkb}, $V_0$ represents the spectral density given by the Cardy density, explained in more detail in Section~\ref{ssec:sur}. For $V$, write
\begin{align}
    V = \sum_{i=1}^6 V_i + V_{\rm torus} + V_{\rm annulus},
\end{align}
where the first six terms arise from six independent bootstrap constraints, and the last two terms come from modular invariance on the torus and the open-closed duality on the annulus. For convenience, we will refer to $\sum_iV_i$ as the tensor potential and $V_{\rm torus}+V_{\rm annulus}$ as the matrix potential, though this nomenclature is not totally accurate due to some part of $V_{\rm annulus}$ having a tensorial nature, as we will see later.

\subsection{Tensor potential}

To form the tensor potential, we square each of the constraints in Section~\ref{sec:cstrs} with an appropriate measure and add them together. 

For the first constraint, square it with the measure $\mu_1(P,\bar{P})$:
\begin{align}
    V_1=c_1\sum_{i,j,k,l} \int_{\mathbb{R}_+^2 \cup \id}\left|\mathcal{M}_{1,ijkl}(P,\bar{P})\right|^2 \mu_1(P,\bar{P}) \,\d P\d\bar{P}.
\end{align}
The notation $|\cdot|^2$ here represents multiplying the quantity by its complex conjugate, but we will use the same notation for multiplication by the anti-holomorphic counterpart, as in the next equation. It should be clear from the context. The overall factor of $c_1$ is a choice, which we will fix later by normalization of the kinetic term. The choice of the measure comes from using the Verlinde inner product on the conformal blocks \cite{Verlinde:1989ua}. (In practice, we can figure out the correct measure by requiring that the second term in \eqref{eq:M1} be squared to a delta function.) Explicitly,
\begin{align}
\mu_1(P,\bar{P})=\frac{1}{|\rho_0(P) C_0(ijP) C_0(kl P)|^2},
\end{align}
where the function $\rho_0(P)$ is the Cardy density:
\begin{align}
    \rho_0(P)=\sinh(2 \pi b P) \sinh ({2 \pi}{b^{-1}} P),
\end{align}
and $C_0(ijk)$, which is a shorthand for  $C_0(P_i,P_j,P_k)$, is the DOZZ formula \cite{Dorn:1994xn,Zamolodchikov:1995aa} up to some normalization \cite{Collier:2019weq}.

Expanding the square and using the following definition of the 6$j$ symbol,
\begin{align}
\left\{\begin{array}{lll}
n & l & i \\
m & j & k
\end{array}\right\}\equiv\frac{C_0(iln) C_0(jkn)}{\rho_0(m)} \mathbb{F}_{nm}\!\begin{bmatrix}
        k & l\\
        j & i
    \end{bmatrix},
\end{align}
the potential $V_1$ simplifies to
\begin{align}\label{eq:V1expanded}
V_1= 2c_1\sum_{i,j,k,l,m,n} \Bigg(&\frac{C_{ijm} C_{mkl} C_{nji} C_{lkn}}{\left|\rho_0(m) C_0(ijm) C_0(klm)\right|^2} \delta^2(P_m-P_n)\nn
-&\frac{C_{ijm} C_{mkl} C_{iln} C_{nkj}}{\left|C_0(ijm) C_0(klm) C_0(jkn) C_0(iln)\right|^2}\left|\left\{\begin{array}{lll}
n & l & i \\
m & j & k
\end{array}\right\}\right|^2\Bigg).
\end{align}

Next, square the second constraint with an appropriate measure (again fixed by the Verlinde inner product):
\begin{align}
V_{2}=c_2\sum_i \int\left|M_{2,i}(P, \bar{P})\right|^2 \mu_{2}(P,\bar{P})\,\d P\d\bar{P},
\end{align}
where
\begin{align}
\mu_2(P,\bar{P})=|\rho_0(P) C_0(iPP)|^2.
\end{align}
Expanding the square leads to
\begin{align}\label{eq:V2expanded}
V_2=2c_2\sum_{i, j, k} \frac{C_{i j j} C_{i k k}}{\left|\rho_0(j) C_0(i j j)\right|^2}\left(\delta^2(P_j-P_k)-\left|\mathbb{S}_{P_jP_k}[P_i]\right|^2\right).
\end{align}

For the third constraint, write
\begin{equation}
    V_3=c_3
    \sum_{a,b,c,d}\sum_{I \in \mathcal{H}^{ab}}
    \sum_{J \in \mathcal{H}^{bc}}
    \sum_{K \in \mathcal{H}^{cd}}
    \sum_{L \in \mathcal{H}^{da}}
    \int_{\mathbb{R}_+ \cup \mathds{1}}\left|\mathcal{M}^{(abcd)}_{3,IJKL}(P)\right|^2 \mu_3(P) {\d P},
\end{equation}
where
\begin{equation}
    \mu_3(P)=\frac{1}{\rho_0(P) C_0(IJP) C_0(KLP)}.
\end{equation}
Expanding the square, it becomes
\begin{align}\label{eq:V3expanded}
    V_3&=2c_3\sum_{a,b,c,d}\sum_{I,J,K,L,M,N}
   \frac{B_{IJM}^{(a b c) } B_{KL M}^{(c d a)}
    B_{NJI}^{(acb) } B_{NLK}^{(cad)}}{g_ag_c}
     \frac{\delta(P_M-P_N)}{\rho_0(M) C_0(IJM) C_0(KLM)} \\
    &-2c_3
    \sum_{a,b,c,d}\sum_{I,J,K,L,M,N}
   \frac{ B_{IJM}^{(a b c) } B_{KL M}^{(c d a)}
   B_{NIL}^{(d ba)} B_{NKJ}^{(bdc)}}{(g_ag_cg_bg_d)^{1/2}}
    \frac{\begin{Bmatrix}
        N & L & I \\
        M & J & K
    \end{Bmatrix}}
    {
    C_0(IJM)C_0(KLM)C_0(ILN)C_0(JKN)
    }.\nonumber
\end{align}

To form the fourth part of the potential, we do the usual thing of squaring, except that here we find it convenient to insert a relative phase of $\e^{\i 2\pi s_i}$ between the two terms in the constraint before squaring (e.g., just add $\e^{\i 2\pi s_i}$ to the second term in $\mathcal{M}^{(ab)}_{4,iIJ}(P)$). We are free to do so because it is trivial for integers $s_i$. It will be useful when we relax the integrality condition. Now
\begin{equation}
    V_4=c_4
    \sum_{a,b}\sum_{i,I,J}
    \int_{\mathbb{R}_+ \cup \mathds{1}}\left|\mathcal{M}^{(ab)}_{4,iIJ}(P)\right|^2 {\mu_4(P)} {\d P},
\end{equation}
where 
\begin{equation}
    \mu_4(P)=
    \frac{1}{\rho_0(P) C_0(i \ib P) C_0(IJP)}.
\end{equation}
Expanding the square, we find
\begin{align}\label{eq:V4expanded}
    V_4
    &=
    \frac{2c_4}{g_a^2}\sum_{a,b}\sum_{i,I,J,M,N}
    \frac{D^{(a)}_{iM}B^{(aab)}_{MIJ} D^{(a)}_{iN}B^{(aab)}_{NJI}\delta(P_M-P_N)}{\rho_0(M) C_0(i\ib M) C_0(IJM)}
    \nn
    &-
    \frac{2c_4}{g_ag_b}
    \sum_{a,b}\sum_{i,I,J,M,N}
    \frac{D^{(a)}_{iM}B^{(aab)}_{MIJ}
    D^{(b)}_{iN}B^{(bba)}_{NIJ} }{{\rho_0(M) C_0(i\ib M) C_0(IJM)}} 
    \nn
    &\qquad\qquad\qquad\qquad\times\int_0^\infty \d P\, 
    \mathbb{F}_{N P}
    \begin{bmatrix}
        I & i \\
        J & \ib
    \end{bmatrix}
    \mathbb{F}_{P M}
    \begin{bmatrix}
        i & \ib \\
        I & J
    \end{bmatrix} 
    \e^{-\i\pi(2h+h_N-h_i-\bar{h}_i-h_I-h_J)},
\end{align}
where $h$ is the weight corresponding to the dummy momentum $P$.

For the fifth constraint, 
\begin{equation}
    V_5=c_5
    \sum_a\sum_{1,2,I}
    \int_{\mathbb{R}^2_+ \cup \mathds{1}}\left|\mathcal{M}^{(a)}_{12I}(P,\bar{P})\right|^2 \mu_5(P,\bar{P}) \,{\d P}\d\bar{P},
\end{equation}
where
\begin{equation}
    \mu_5(P,\bar{P})=\frac{1}{\rho_0(P)\rho_0(\bar{P})C_0(ijP)C_0(\ib\jb\bar{P})C_0(IP\bar{P})}.
\end{equation}
Expanding the square, this becomes
\begin{align}\label{eq:V5expanded}
    V_5
    &=
    {c_5}
    \sum_a\sum_{i,j,k,l,I} C_{ijk} C_{lji} D_{kI}^{(a)} D_{lI}^{(a)}  \frac{\delta^{2}(P_k-P_l)}{\rho_0(k)\rho_0(\bar{k})C_0(ijk)C_0(\ib \jb \bar{k})C_0(Ik\bar{k})}
    \\
    &+c_5\sum_a\frac{1}{g_a^4}\sum_{i,j,I,J,K,L,M} B_{MIJ}^{(aaa)} B_{LKM}^{(aaa)} D_{iI}^{(a)} D_{jJ}^{(a)} D_{iK}^{(a)} D_{jL}^{(a)} \frac{\delta(P_I-P_K)\delta(P_J-P_L)}{\rho_0(I)\rho_0(J)C_0(i\ib I)C_0(j\jb J)C_0(MIJ)}
    \nn
    &- c_5\sum_a\Bigg[\frac{1}{g_a^2} \sum_{i,j,k,I,J,K} B_{IJK}^{(aaa)} C_{kji} D_{iI}^{(a)} D_{jJ}^{(a)} D_{kK}^{(a)} \frac{\mathbb{K}[P_I,P_J;P_k,\bar{P}_k; P_K,P_i,P_j]}{\rho_0(k)\rho_0(\bar{k})C_0(ijk)C_0(\ib\jb\bar{k})C_0(k\bar{k}K)}+\cc \Bigg].\nonumber
\end{align}

Finally, for the sixth constraint,
\begin{equation}
     V_6=c_6
     \sum_{a,b}\sum\limits_{\substack{I\in\mathcal{H}^{aa} \\ J\in\mathcal{H}^{bb}}}
    \int_{\mathbb{R}^2_+ \cup \mathds{1}}\left|\mathcal{M}^{(ab)}_{6,IJ}(P,\bar{P})\right|^2 \mu_6(P,\bar{P}) {\d P}\d\bar{P},
\end{equation}
where
\begin{equation}
    \mu_6(P,\bar{P})=\frac{1}{\rho_0({P})\rho_0(\bar{P})C_0(IP\bar{P})C_0(JP\bar{P})}.
\end{equation}
Expanding the square, we get
\begin{align}\label{eq:V6expanded}
    V_6&=c_6\sum_{a,b}\frac{1}{g_ag_b}\sum_{i,j,I,J} D_{iI}^{(a)} D_{iJ}^{(b)} D_{jI}^{(a)} D_{jJ}^{(b)} \mu_6(P_i,\bar{P}_i)\delta^{2}(P_i-P_j) \\
    &+ c_6\sum_{a,b}\frac{1}{(g_a g_b)^3}\sum_{I,J,K,L,M,N} B_{IKL}^{(aab)} B_{JKL}^{(bba)} B_{NMI}^{(aba)} B_{NMJ}^{(bab)}\frac{\delta(P_L-P_N)\delta(P_K-P_M)}{\rho_0(K) \rho_0(K)C_0(IKK)C_0(JKK)}\nn
    &- c_6\sum_{a,b}\Bigg[\frac{1}{(g_a g_b)^2}\sum_{i,I,J,K,L} D_{iI}^{(a)} D_{iJ}^{(b)} B_{IKL}^{(aab)} B_{JKL}^{(bba)} 
    \frac{\mathbb{K}_{6} [P_K,P_L;P_i, \bar{P}_i;P_I, P_J] }{{\rho_0({i})\rho_0(\ib)C_0(i\ib I)C_0(i\ib J)}} +\cc\Bigg].\nonumber
\end{align}

\subsection{Matrix potential}

To form the matrix potential, \cite{Jafferis:2024jkb} used the modular invariance of the torus partition function without insertions (empty torus), which is given by \eqref{eq:cstr2} with $i=\id$:
\begin{equation}
    \sum_{j}\delta^{2}(P-P_j) - \sum_j |\mathbb{S}_{j P}[\id]|^2=0.
\end{equation}
Unlike $V_1$ to $V_6$, the blocks associated with the empty torus are not normalizable with respect to the Verlinde inner product. For this reason, \cite{Jafferis:2024jkb} proposed using the Vandermonde measure: 
\begin{align}
    K(\Delta, s ; \Delta', s')=\delta_{s s'} \log \left|\Delta-\Delta'\right|.
\end{align}
A matrix potential $V_{\rm torus}$ is formed by squaring the empty torus constraint with this measure. Its explicit expression and associated subtleties related to the continuity of spin can be found in \cite{Jafferis:2024jkb}.

We now do the analogous thing for the empty annulus.\footnote{See \cite{Post:2024itb} for a discussion of the boundary one-point function on the annulus, which is somewhere in between the annulus two-point function where we use the Verlinde inner product and the empty annulus where we use the Vandermonde.} Taking the annulus constraint \eqref{eq:ctsr6} and setting $I=J=\id$, the constraint becomes
\begin{align}
    \sum_{m} A_{m}^{(a)} A_{m}^{{(b)}} \delta^{2}(P-P_m)  -\sum_{M} \mathbb{S}_{ M P}[\mathbbm{1}] \delta(P-\bar{P}),
\end{align}
where we have used \cite{Numasawa:2022cni}
\begin{align}
    \mathbb{K}_{6}[M,M; P, \bar{P};\mathbbm{1}, \mathbbm{1}] 
    &
    = \int {\d P'} 
    \mathbb{F}_{ M P'} \begin{bmatrix}
        M & M \\
        \mathbbm{1} & \mathbbm{1}
    \end{bmatrix} 
    \mathbb{S}_{ M P}[P'] \mathbb{F}_{ P' \bar{P}} \begin{bmatrix}
        P & \mathbbm{1} \\
        P & \mathbbm{1}
    \end{bmatrix}
    \nn
    &=
    \mathbb{S}_{ M P}[\mathbbm{1}] \delta(P-\bar{P}).
\end{align}
We now need to square this with an appropriate measure. However, it is more convenient to assign the measure if we write the same constraint by expressing the open channel in terms of the closed channel:
\begin{equation}\label{eq:cstr_ann}
    \sum_{I\in\mathcal{H}^{ab}}\delta(P-P_I) - \sum_i A_i^{(a)} A_i^{(b)} \mathbb{S}_{i P}[\id]=0.
\end{equation}

Analogous to the closed sector, the measure on the open-sector weights is given by the Vandermonde:
\begin{align}
    K(h, h') =  \log \left|h - h'\right| \, .
\end{align}
Define $\tilde{K}$ via
\begin{align}
    \int \d h \d h' K(h,h') [\cdot] =& \int (P \d P)( P'\d P') K(h(P),h'(P')) [\cdot]
    \nn
    \equiv& \int \d P\d P'\tilde{K}(P,P')[\cdot],
\end{align}
where $[\cdot]$ represents some generic integrand.

Squaring the constraint \eqref{eq:cstr_ann} and integrating over $P, P'$ using the measure $\tilde{K}(P,P')$, we obtain:
\begin{equation}\label{eq:Vann}
\begin{aligned}
    V_{\rm annulus}&=\sum_{a,b}\sum_{I,J\in\mathcal{H}^{ab}} {\tilde{K}(P_I,P_J)} - 2\sum_{a,b}\sum_{i,I\in\mathcal{H}^{ab}} A_{i}^{(a)} A_{i}^{(b)} \int {\d P \tilde{K}}(P,P_I) \mathbb{S}_{i P}[\mathbbm{1}]\nn
    &+\sum_{a,b} \sum_{i,j} A_{i}^{(a)} A_{i}^{(b)} A_{j}^{(a)} A_{j}^{(b)} \int {\d P \d P'}  \mathbb{S}_{i P}[\mathbbm{1}] {\tilde{K}}(P,P') \mathbb{S}_{P' j}[\mathbbm{1}].
\end{aligned}
\end{equation}We have used that $\mathbb{S}_{ij}[\mathbbm{1}]$ is real, so complex conjugating does nothing. 

We propose that the matrix ensemble is of GOE type when $a=b$ and of GUE type when $a\ne b$. To see this, consider a strip with boundary conditions $a$ and $b$ on the two sides. If $a=b$, CRT maps the strip to itself, but not if $a\ne b$. The difference is a factor of two. Interestingly, this difference is already captured by the matrix potential \eqref{eq:Vann}. When $a\ne b$, there is a factor of two compared to when $a=b$ because $\mathcal{H}^{ab}=\mathcal{H}^{ba}$. We will see how 3d gravity realizes this factor of two in Section~\ref{ssec:offshell}. See e.g.~\cite{Yan:2023rjh} for relevant discussions in the closed case. 

It is useful to separate out the contributions where bulk operator indices ($i$,$j$) are equal to the identity. The expression then turns into
\begin{align}\label{eq:Vann_long}
    V_{\rm annulus}&=\sum_{a,b}\sum_{I,J} {\tilde{K}(P_I,P_J)} 
    - 2\sum_{a,b}\sum_{I} g_a g_b \int {\d P \tilde{K}}(P,P_I) \mathbb{S}_{\mathbbm{1} P}[\mathbbm{1}]
    \nn
    &- 2\sum_{a,b}\sum_{i}'\sum_{I} A_{i}^{(a)} A_{i}^{(b)} \int {\d P \tilde{K}}(P,P_I) \mathbb{S}_{P_i P}[\mathbbm{1}]\nn
    &+2\sum_{a,b}\sum_{i}' A_{i}^{(a)} A_{i}^{(b)} g_a g_b \int {\d P \d P'}  \mathbb{S}_{P_i P}[\mathbbm{1}] {\tilde{K}}(P,P') \mathbb{S}_{P' \mathbbm{1}}[\mathbbm{1}]
    \nn
    &+\sum_{a,b}\sum_{i,j}' A_{i}^{(a)} A_{i}^{(b)} A_{j}^{(a)} A_{j}^{(b)} \int {\d P \d P'}  \mathbb{S}_{P_i P}[\mathbbm{1}] {\tilde{K}}(P,P') \mathbb{S}_{P' P_j}[\mathbbm{1}],
\end{align}
where we have used $A^{(a)}_\id=g_a$ and dropped a term
\begin{equation}
    \sum_{a,b} g_a^2 g_b^2 \int {\d P \d P'}  \mathbb{S}_{\mathbbm{1}P}[\mathbbm{1}] {\tilde{K}}(P,P') \mathbb{S}_{P' \mathbbm{1}}[\mathbbm{1}],
\end{equation}
which is just a number that shifts the potential by a constant. 

Now, the first term of \eqref{eq:Vann_long} is the kinetic term of the matrix model, and its functional inverse gives the matrix two-point function $\langle \rho^{ab}(P_I)\rho^{ab}(P_J)\rangle$ where $\rho^{ab}$ denotes the density of states in $\mathcal{H}^{ab}$. The factor of two in the potential for $a\ne b$ becomes a factor of half in the density two-point function, reflecting the nature of the ensemble. The second term of \eqref{eq:Vann_long} is a single-trace term. The other terms involve the dynamical objects $A$'s, so the matrices interact non-trivially with these vectors. We will discuss the diagrams associated with these expressions in the next section.

\section{3d interpretations}\label{sec:3d}
The tensor-matrix model is a quantum mechanical model. One way to study it is the perturbative approach of Feynman diagrams. Interestingly, in \cite{Jafferis:2024jkb}, the Feynman rules of the closed tensor-matrix model map to 3d diagrams with the associated functions mapped to partition functions of 3d gravity. An interesting question we can now ask is: What 3d manifolds do Feynman diagrams of our new tensor-matrix model map to? In other words, what is the 3d theory whose partition functions on those manifolds give us the correct expressions that reproduce the Feynman rules we have?

In this section, we identify the 3d ingredients one by one. We start with the dynamical variables and identify their 3d counterparts. We then move on to propagators and vertices. We will find that these give rise to an open-closed extension of Virasoro TQFT. At the end, we will also study the 3d avatars of the matrix diagrams which for example give predictions for partition functions on off-shell topologies such as the annulus wormhole, the open analog of the torus wormhole.

\subsection{Data}
The OPE coefficients $C$, $B$ and $D$ capture the information contained in the three-point function of bulk operators on the sphere, the three-point function of boundary operators on the disk, and the one-bulk-one-boundary two-point function on the disk, respectively, so they are associated with the pictures \eqref{eq:datapics}, reproduced here for convenience:
\begin{align}
   C_{ijk} \longrightarrow \vcenter{\hbox{\includegraphics[height=1.5cm]{figs/dataC.pdf}}}~,
   \qquad B_{IJK}^{(abc)} \longrightarrow \vcenter{\hbox{\includegraphics[height=1.5cm]{figs/dataB.pdf}}}~,
   \qquad D_{iI}^{(a)} \longrightarrow \vcenter{\hbox{\includegraphics[height=1.5cm]{figs/dataD.pdf}}}~.
\end{align}
These 2d pictures appear as cross-sections of the 3d diagrams. Notice the small arrows in the representation of $C_{ijk}$ and $D_{iI}^{(a)}$. We have included them in anticipation of the framing that will be of importance from the TQFT point of view. At each cross-section where a bulk Wilson line passes through, the framing of the Wilson line is represented by a little arrow. The boundary operators do not carry arrows, as the boundary Wilson lines do not have framing. We will often suppress the drawing of the framing when it is not the main focus. 

The corresponding diagram for the final piece of OPE data $A_i^{(a)}$ is not shown, but it is the same as $D_{iI}^{(a)}$ with the boundary insertion removed. For simplicity, we will often not draw the diagrams involving $A$'s separately when we can obtain them from the diagrams involving $D$'s.

\subsection{Propagators}
To find the propagators, we now single out the terms in the potential that correspond to the kinetic terms for $C$, $B$, and $D$, respectively. 

\subsubsection*{\texorpdfstring{$C$}{C}-propagator}
For $C_{ijk}$, the kinetic term can be obtained by identifying the quadratic terms in the second line of \eqref{eq:V3expanded} \cite{Jafferis:2024jkb}. The result is
\begin{equation}\label{eq:Ckin}
K_C=- 4c_1\sum_{i, j,k}' \frac{C_{ijk} C_{kji}}{\left|C_0(ijk)\right|^2}.
\end{equation}
The factor of four comes from setting different labels to the identity. It is therefore natural to pick $c_1=1/4$. Inverting this gives the propagator, which we associate with the diagram
\begin{align}\label{eq:C0}
   -{\hbar} \, |C_0(ijk)|^2 \longrightarrow \vcenter{\hbox{\includegraphics[height=1.5cm]{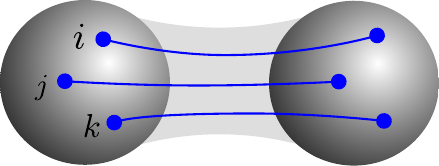}}}~.
\end{align}

\subsubsection*{\texorpdfstring{$B$}{B}-propagator}
The kinetic term for $B$ comes  from $V_3$ \eqref{eq:V3expanded}. When one of the internal operators, say $N=\mathds{1}\in\mathcal{H}^{bd}$ (so $d=b$), is taken to be the identity, we must have $I\in\mathcal{H}^{ab}=L\in\mathcal{H}^{da}$ and $J\in\mathcal{H}^{bc}=K\in\mathcal{H}^{cd}$. This gives
\begin{align}
    K_B=-4c_3 \sum_{I, J,K}'  \frac{B_{IJK}^{(abc)} B_{JIK}^{(cba)} }{C_0(IJK)},
\end{align}
where we used
\begin{align}
    {B}_{II\mathds{1}}^{(bab) } {B}_{JJ \mathds{1}}^{(bcb)}
    &=\sqrt{g_{a}g_b}\sqrt{g_{b}g_c};
    \\
    \begin{Bmatrix}
        \mathds{1} & I & I \\
        M & J & J
        \end{Bmatrix}
        &=\rho_0(M)^{-1}C_0(II\mathds{1})C_0(JJ\mathds{1})\mathbb{F}_{\mathds{1}M}\begin{bmatrix}
    J & I \\
    J & I
    \end{bmatrix};
    \\
    \mathbb{F}_{\mathds{1}M}\begin{bmatrix}
    J & I \\
    J & I
    \end{bmatrix}
    &=C_0(IJM)\rho_0(M).
\end{align}
The extra factor of $g_b$ comes from removing the $N$-sum which contains the inverse metric $\mathfrak{g}^{(ab)IJ}=(g_ag_b)^{-1/2}\delta^{IJ}$. The factor of two comes from the fact that we can set either $M$ or $N$ to the identity. Also, if both $M$ and $N$ are set to the identity, it gives a constant to the potential which is then dropped. Incidentally, the first line in \eqref{eq:V3expanded} with $I=J,K=L,N=\mathds{1}$ but $M\ne\mathds{1}$ is zero because of the delta function, so it does not contribute to the kinetic term.

Picking $c_3=1/4$, we now invert the kinetic term to obtain the propagator and assign to it the following 3d diagram:
\begin{align}\label{eq:propB}
   -{\hbar} \, C_0(IJK) \longrightarrow \vcenter{\hbox{\includegraphics[height=1.5cm]{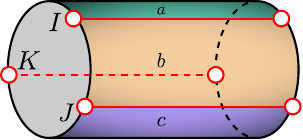}}}~.
\end{align}

\subsubsection*{\texorpdfstring{$D$}{D}-propagator}
We now turn to $D$. From the third term of $V_5$ \eqref{eq:V5expanded}, we have, upon setting $M=N\ne\id$, $i=j$, $I=\mathbbm{1}$, $k=\mathbbm{1}$,
\begin{equation}
    -\Bigg[ \frac{c_5}{g_a^3} \sum'_{i}\sum_{M} (B^{(aaa)}_{MM\mathbbm{1}} C_{ii\id}^*D_{\mathbbm{1}\mathbbm{1}}^{*(a)})
    D_{iM}^{(a)} D_{iM}^{(a)}  \frac{\mathbb{K}[M,M;\mathbbm{1},{\mathbbm{1}}; \mathbbm{1},i,i]}{\rho_0(\mathbbm{1})\rho_0({\mathbbm{1}})C_0(ii\mathbbm{1})C_0(\ib \ib {\mathbbm{1}})C_0(\mathbbm{1}\mathbbm{1}{\mathbbm{1}})} +\cc\Bigg].
\end{equation}
Using (see e.g.~\cite{Eberhardt:2023mrq})
\begin{align}
    \mathbb{K}[M,M;\id,{\mathbbm{1}}; \mathbbm{1},i,i] &\equiv \e^{\i\pi \bar{h}_i}\int_0^{\infty} \d P_r\,  \e^{-\i\pi h_r} 
    \mathbb{F}_{M r} \begin{bmatrix}
        \mathbbm{1} & \ib \\ M & i 
    \end{bmatrix} 
    \mathbb{F}_{M \mathbbm{1}} \begin{bmatrix}
        \ib & r \\ i & i
    \end{bmatrix} 
    \mathbb{F}_{r \ib} \begin{bmatrix}
        \ib & \ib \\ 
        \mathbbm{1} & \mathbbm{1}
    \end{bmatrix}
    \nn
    &=
    \frac{\rho_0(\mathbbm{1})C_0(ii\mathbbm{1} )C_0(\ib\ib\mathbbm{1} )}
    {C_0(i\ib M)}
    \times \rho_0(\mathbbm{1})C_0(\mathbbm{1}\mathbbm{1}\mathbbm{1} ),
\end{align}
we have
\begin{align}\label{eq:Dkin}
    K_D
    =- 2c_5 \sum'_{i,I} 
    \frac{D_{iI}^{(a)} D_{iI}^{*(a)}}
    {C_0(i\ib I)}, 
\end{align}
where we have used $B^{(bab)}_{MM\mathbbm{1}}=\sqrt{g_bg_a}$ and $D_{\mathbbm{1}\mathbbm{1}}=g_a$. Picking $c_5=1/2$ and inverting the kinetic term, we obtain the propagator
\begin{align}
   -\hbar \,C_0(i\ib I) \longrightarrow \vcenter{\hbox{\includegraphics[height=1.5cm]{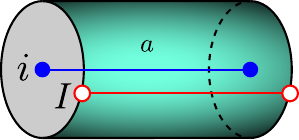}}}~.
\end{align}

\subsubsection*{\texorpdfstring{$A$}{A}-propagator}
The derivation for $A$ is essentially the same as that for $D$. From the third term of $V_5$ \eqref{eq:V5expanded}, upon setting $M=N=\id$, $i=j$, $I=\mathbbm{1}$, $k=\mathbbm{1}$, we get
\begin{align}
    K_A
    =-2c_5 \sum'_{i} 
      \frac{A_{i}^{(a)} A_{i}^{(a)}}
    {\delta(P_i-\bar{P}_i)}\rho_0(i), 
\end{align}
where we have used $C_0(i\ib \mathbbm{1})=\delta(P_i-\bar{P}_i)/\rho_0(i)$. Therefore, we make the association
\begin{align}
   -\hbar\frac{\delta(P_i-\bar{P}_i)}{\rho_0(i)} \longrightarrow \vcenter{\hbox{\includegraphics[height=1.5cm]{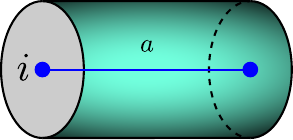}}}~.
\end{align}
The delta function shows that there is no propagator unless the operator labeled by $i$ is a scalar. These Wilson lines therefore have no framing.

It is interesting to point out that there is no classical solution with this topology where the punctured disks are asymptotic boundaries and the bulk Wilson line is below the black hole threshold, unlike the other three propagators. This is most easily seen by gluing two such geometries along the brane to form a diagram with the topology of a twice-punctured sphere times an interval. This is not an issue because the states represented by the Wilson line only have support above the black hole threshold, so we should not look for classical solutions with conical defects replacing Wilson lines. If we sum over the index $i$, we should replace the Wilson line with a tube connecting the two asymptotic disks, turning the asymptotic geometry into a single annulus. In that case, there is indeed an on-shell solution with that topology \cite{Takayanagi:2011zk}. The procedure of turning a Wilson line connecting asymptotic boundaries into a tube is explained in \cite{Jafferis:2024jkb} and reviewed in Section~\ref{ssec:sur}, where an analogous procedure for boundary Wilson lines is also explained.

\subsection{Open-closed TQFT}\label{ssec:ocTQFT}

An open-closed TQFT is a TQFT where two manifolds can be glued along certain open subsets of their boundaries \cite{Lazaroiu:2000rk,Lauda:2005wn,Moore:2006dw}. For example, in a 3d closed TQFT, two three-balls must be glued along their entire boundaries, which are two-spheres, resulting in a three-sphere. In contrast, in a 3d open-closed TQFT, two three-balls can be glued along a disk, which is a subset of each of their boundaries, resulting in a three-ball. In this section, we explain how the tensor part of the model constructed in the previous section leads to an open-closed Virasoro TQFT. In the rational case, the relation between 3d TQFT and 2d BCFT bootstrap has been extensively studied in \cite{Felder:1999cv,Felder:1999mq,Fuchs:2002cm,Fuchs:2003id,Fuchs:2004dz,Fuchs:2004xi,Fjelstad:2005ua,Frohlich:2006ch,Fjelstad:2006aw,Kong:2009inh,Kong:2013gca,Traube:2020pfg}. Our language will be most closely following \cite{Collier:2023fwi}, and we emphasize the open-closed nature of our construction, in contrast to e.g.~\cite{Kong:2013gca}, which constructs a closed TQFT using the doubling trick. In our case, since the topology of the EOW brane is essential in reproducing the correct $g$-function dependence, the open sector is non-trivial, meaning that it is not equivalent to a doubled version.

To obtain TQFT diagrams, we should first turn the 2d diagrams in equations like \eqref{eq:cstr1block} into 3d diagrams. Let us discuss \eqref{eq:cstr1block} first. The procedure is simple. Think of the sphere as the boundary of a 3-ball, and imagine continuously deforming the skeleton into the interior of the ball whilst keeping the four insertions anchored at the boundary (the 2-sphere). Now, taking the diagonal terms is tantamount to gluing two of the same-channel block diagrams (now 3d) along their boundaries (each a 2-sphere with 4 insertions) with the insertions appropriately identified ($i$ with $i$, $j$ with $j$, and so on). The result is a skeleton diagram in $S^3$. We can then remove a 3-ball in the neighborhood of each trivalent junction. This leads to the following manifold:
\begin{align}\label{eq:pillow}
   \frac{2c_1}{\hbar}\frac{\delta^2(P_m-P_n)}{\left|\rho_0(m) C_0(ijm) C_0(klm)\right|^2}  &\longrightarrow \vcenter{\hbox{\includegraphics[height=4cm]{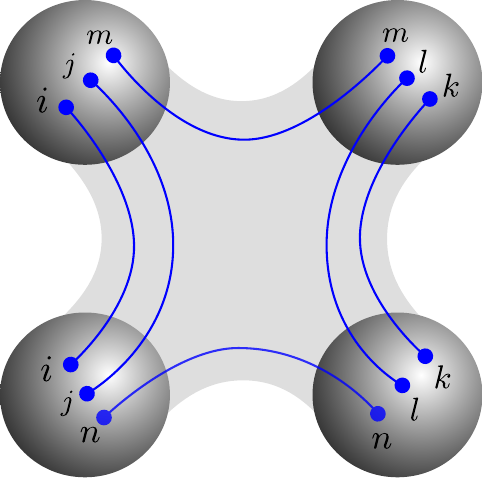}}}~,
\end{align}
with index structure $C_{ijm} C_{mkl} C_{nji} C_{lkn}$ and a coupling of ${2c_1}/{\hbar}$. This was referred to as the pillow diagram in \cite{Jafferis:2024jkb}, but in anticipation of a similar object that will show up soon, we will call this more precisely the closed pillow diagram. It is a 3d manifold with four boundaries, each a sphere thrice punctured.

The cross terms are obtained similarly. Gluing an $s$-channel block to a $t$-channel block leads to a skeleton diagram that forms the edges of a tetrahedron in $S^3$, which becomes the following 3d manifold upon removing a ball at each of the tetrahedron's corners:
\begin{align}\label{eq:6jbulk}
    -\frac{2c_1}{\hbar}\frac{\left|\left\{\begin{array}{lll}
    n & l & i \\
    m & j & k
    \end{array}\right\}\right|^2}{\left|C_0(ijm) C_0(klm) C_0(jkn) C_0(iln)\right|^2}
   &\longrightarrow \vcenter{\hbox{\includegraphics[height=4cm]{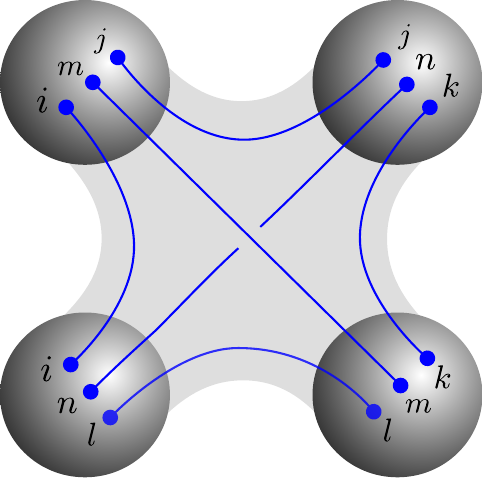}}}~,
\end{align}
with index structure $C_{ijm} C_{mkl} C_{iln} C_{nkj}$ and a coupling of $-{2c_1}/{\hbar}$. We refer to this as the closed 6$j$ manifold to distinguish it from the open 6$j$ manifold appearing later. 

As for $V_2$ presented in \eqref{eq:V2expanded}, the first term comes from the two diagonal terms in expanding the square, and it is associated with the following diagram:
\begin{align}
   \frac{2c_2}{\hbar}\frac{\delta^2(P_j-P_k)}{\left|\rho_0(j) C_0(i j j)\right|^2} &\longrightarrow \vcenter{\hbox{\includegraphics[height=1.5cm]{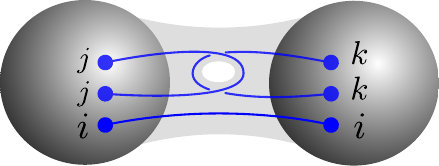}}}~,
\end{align}
with index structure $C_{i j j} C_{i k k}$. The hole in the middle represents a genus, in the sense that the manifold has topology $S^2\times S^1$ when the Wilson lines are stripped and the two boundary spheres are shrunk to points. The second term is assigned the following diagram:
\begin{align}
    -\frac{2c_2}{\hbar}\frac{C_{i j j} C_{i k k}}{\left|\rho_0(j) C_0(i j j)\right|^2}
    \left|\mathbb{S}_{P_jP_k}[P_i]\right|^2
   &\longrightarrow \vcenter{\hbox{\includegraphics[height=1.5cm]{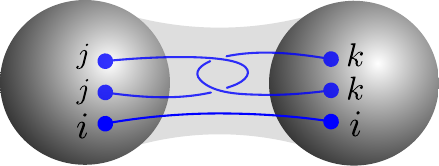}}}~.
\end{align}
Here, there is no genus hole in the bulk of the manifold, but the Wilson lines labeled by $j$ and $k$ are interlocked to prevent them from deforming continuously to lie on the boundaries.

To obtain these diagrams, we again uplift the 2d diagrams in \eqref{eq:cstr2block} to 3d diagrams. For the first diagram, we continuously deform the blue skeleton into the interior of the solid torus (as drawn), keeping the insertion anchored at the torus. For the second diagram, we should continuously deform the skeleton outward and take the exterior of the solid torus (as drawn) to be the 3d diagram of interest. Gluing two identical blocks therefore gives a 3d manifold with topology $S^2\times S^1$, while gluing two different blocks gives a 3-manifold with topology $S^3$. The way the two solid tori are glued together to become $S^3$ is an example of Heegaard splitting. The Wilson line structure is obtained in the same way as before, though in this case we only need to remove two 3-balls, resulting in diagrams that each have two boundary spheres. 

Now let us move on to the new terms in the potential. The procedure explained so far does not work anymore because the 2d surfaces now have boundaries. 

We now describe a generalization of the procedure used in the closed sector. To begin with, notice that unlike for the first two constraints which are concerned with correlators on closed Riemann surfaces, we now have correlators on the disk and the annulus. A disk is not the boundary of any 3d manifold, so we need some extra ingredients to circumvent this issue. The idea is to join the disk with an EOW brane with disk topology to obtain a surface with sphere topology and fill in the sphere with a 3-ball. This is a familiar idea from the AdS/BCFT proposal. We then proceed as before by deforming the red skeleton into the interior of the 3d manifold with the red circles anchored. However, unlike for blue skeletons, we need to keep deforming it until it entirely lies on the EOW brane, as below:
\begin{align}
    \vcenter{\hbox{\includegraphics[scale=0.6]{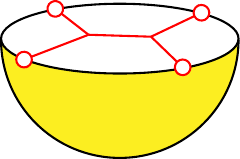}}} \longrightarrow \vcenter{\hbox{\includegraphics[scale=0.6]{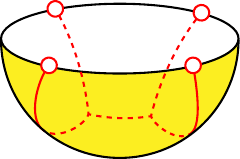}}}~.
\end{align}
We then glue two of these 3d manifolds along the disks to form a 3-ball whose boundary is an EOW brane decorated with a red skeleton, the same boundary operators being glued together. We do not glue the EOW branes because they are not part of the geometry describing the BCFT correlator. For example, when we glue an $s$-channel diagram to a $t$-channel one, we get
\begin{align}
    \vcenter{\hbox{\includegraphics[scale=0.6]{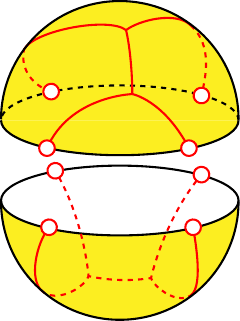}}} \longrightarrow \vcenter{\hbox{\includegraphics[scale=0.6]{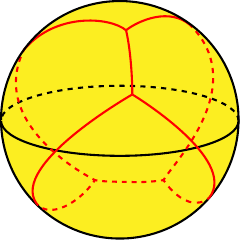}}}~.
\end{align}
It is not hard to realize that the upper hemisphere needs to be the mirror image of the $t$-channel diagram (the lower hemisphere being the $s$-channel diagram). Because the disk on the lower diagram has $I,J,K,L$ in a clockwise order (as viewed from below), the disk on the upper diagram must have the opposite orientation in order to be glued. In terms of formulae, taking the mirror image corresponds to complex conjugation, as suggested by \eqref{eq:RP} and \eqref{eq:datapics}.

Consider $V_3$ in \eqref{eq:V3cross}. The first term is the open-sector analog of the closed pillow diagram in \eqref{eq:pillow}, which we refer to as the boundary pillow diagram:
\begin{align}\label{eq:V3diag}
   \frac{2c_3}{\hbar}\frac{\delta(P_M-P_N)}{g_ag_c\rho_0(M) C_0(IJM) C_0(KLM)}
    &\longrightarrow \vcenter{\hbox{\includegraphics[height=3.5cm]{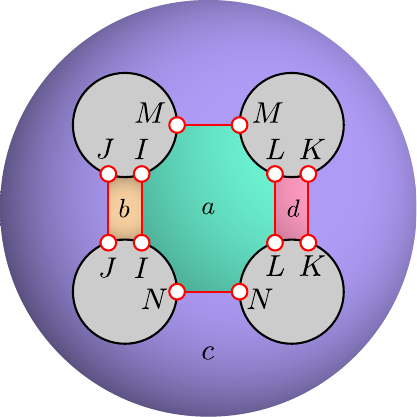}}}~,
\end{align}
with index structure $B_{IJM}^{(a b c) } B_{KL M}^{(c d a)}    B_{NJI}^{(acb) } B_{NLK}^{(cad)}$.
In this diagram, four disks are connected via a 3d manifold accompanied by a collection of EOW branes. Each brane has a ``color" label, $a$. Two branes with labels $a$ and $b$ can meet at what we will call a boundary Wilson line, each representing a state in the open-string Hilbert space $\mathcal{H}^{ab}$. The branes can also end on intervals at the boundary of disks representing $B$ (and also $D$ as we will see later). The label of the brane must match the label of the boundary condition when it does so.

As for the second term in the potential $V_3$ in \eqref{eq:V3expanded}, we associate it with the diagram 
\begin{align}\label{eq:V3cross}
    -\frac{2c_3}{\hbar}\frac{({g_ag_cg_bg_d})^{-1/2}\begin{Bmatrix}
        N & L & I \\
        M & J & K
    \end{Bmatrix}}{C_0(IJM)C_0(KLM)C_0(ILN)C_0(JKN)}
   &\longrightarrow \vcenter{\hbox{\includegraphics[height=3.5cm]{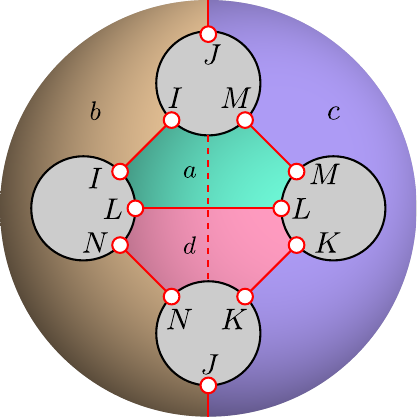}}}~,
\end{align}
with index structure $B_{IJM}^{(a b c) } B_{KL M}^{(c d a)} B_{NIL}^{(d ba)} B_{NKJ}^{(bdc)}$. This is the open-sector analog of the closed 6$j$ manifold, so we will refer to it as the open 6$j$ manifold.\footnote{This is exactly the generalized tetrahedron in the heavy regime, where we glue along the truncated triangles (disks with three boundary operators). See Section~\ref{sec:pureopen} for its relation to the tetrahedral decomposition where the gluing is along the faces. See also \cite{Hartman:2025cyj,Hartman:2025ula}.}  If we glue two such diagrams along the branes, we obtain a diagram that has the topology of the closed 6$j$ manifold.

Next, analogous to the step taken when dealing with the first two constraints, where we remove a ball at each trivalent junction of blue lines (bulk Wilson lines), we remove a half-ball neighborhood for each trivalent junction of red lines. This leaves a disk with three boundary Wilson lines ending on it, and the three intervals must be assigned boundary condition labels in a way consistent with the labels of the boundary Wilson lines. Finally, assign each piece of the EOW brane separated by the boundary Wilson lines a color label so that all intervals in contact with it have the same label for the boundary condition. This procedure leads to the diagrams \eqref{eq:V3diag} and \eqref{eq:V3cross}.

Consider now $V_4$. To proceed, again consider the 2d diagrams for the conformal blocks in \eqref{eq:cstr4block} and upgrade them to 3d diagrams. As before, continuously deform the skeleton into the interior of the 3d manifold bounded by an EOW brane and the disk. The novelty here is that our skeleton now consists of both blue and red parts, but the rules are the same: The blue lines are deformed into the bulk of the 3d manifold, while the red lines must be deformed until they lie entirely on the EOW brane. In this case, we have
\begin{align}
    \vcenter{\hbox{\includegraphics[height=2cm]{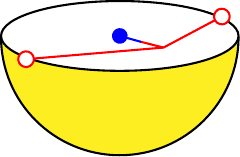}}} \quad\longrightarrow \quad
    \vcenter{\hbox{\includegraphics[height=2cm]{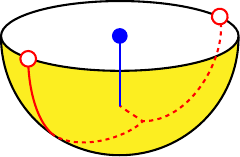}}}~.
\end{align}
Here the blue line extends from the bulk operator insertion on the disk (blue dot) to somewhere on the EOW brane through the 3d interior. The red lines all lie entirely on the EOW brane. Finally, we glue two of such diagrams to obtain a 3-ball whose boundary is an EOW brane with sphere topology. As before, a blue trivalent junction is replaced with a ball having three bulk operator insertions and a red trivalent junction is replaced with a disk having three boundary operator insertions. As for the blue-red junction, which is new, we replace it with the last diagram in \eqref{eq:datapics}, i.e., a disk with a bulk operator and a boundary operator, which represents the bulk-to-boundary OPE. This leads to the following diagrams upon coloring the different pieces of the brane separated by boundary Wilson lines with different colors. For the first term, we have the following vertex for $D^{(a)}_{iK}B^{(aab)}_{KIJ} D^{(a)}_{iL}B^{(aab)}_{LJI}$:
\begin{align}\label{eq:V4diag}
   \frac{2c_4}{\hbar}\frac{\delta(P_K-P_L)}{g_a^2\rho_0(K) C_0(i\ib K) C_0(IJK)}
    &\longrightarrow \vcenter{\hbox{\includegraphics[height=3.5cm]{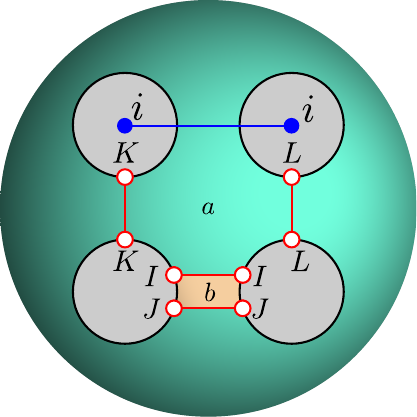}}}~.
\end{align}
For the second term, we have the following vertex for $D^{(a)}_{iK}B^{(aab)}_{KIJ}
    D^{(b)}_{iL}B^{(bba)}_{LIJ}$:
\begin{align}\label{eq:V4cross}
    -\frac{2c_4}{\hbar}\frac{(g_ag_b)^{-1}}{{\rho_0(K) C_0(i\ib K) C_0(IJK)}} &\int_0^\infty \d P\, 
    \mathbb{F}_{L P}
    \begin{bmatrix}
        I & i \\
        J & \ib
    \end{bmatrix}
    \mathbb{F}_{P K}
    \begin{bmatrix}
        i & \ib \\
        I & J
    \end{bmatrix}
    \e^{-\i\pi(2h+h_L-h_i-\bar{h}_i-h_I-h_J)}
   \nn
   &\longrightarrow
   \vcenter{\hbox{\includegraphics[height=3.5cm]{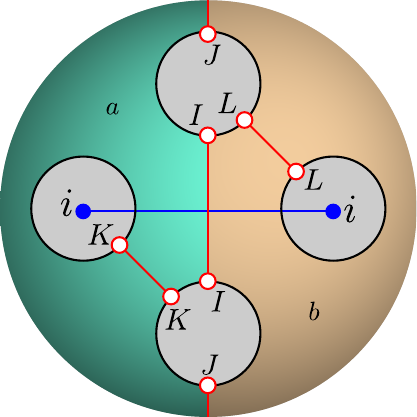}}}~.
\end{align}

For $V_5$, the algorithmic procedure leads to the following diagrams. The first term gives the following vertex for $C_{ijk} C_{lji} D_{kI}^{(a)} D_{lI}^{(a)}$:
\begin{align}\label{eq:V5diag1}
    \frac{c_5}{\hbar}\frac{ \delta^{2}(P_k-P_l)}{\rho_0(k)\rho_0(\bar{k})C_0(ijk)C_0(\ib\jb \bar{k})C_0(Ik\bar{k})}
    &\longrightarrow \vcenter{\hbox{\includegraphics[height=3.5cm]{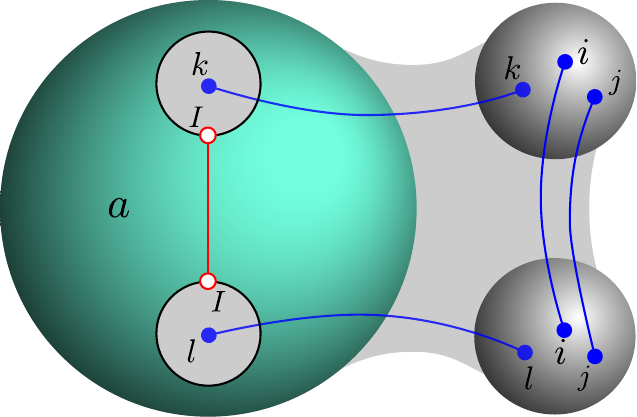}}}~.
\end{align}
The second term gives the following diagram for the vertex $B_{MIJ}^{(aaa)} B_{LKM}^{(aaa)} D_{iI}^{(a)} D_{jJ}^{(a)} D_{iK}^{(a)} D_{jL}^{(a)}$:
\begin{align}
    \frac{c_5}{\hbar}\frac{g_a^{-4} \delta(P_I-P_K)\delta(P_J-P_L)}{\rho_0(I)\rho_0(J)C_0(i\ib I)C_0(j\jb J)C_0(MIJ)}
   &\longrightarrow
   \vcenter{\hbox{\includegraphics[height=3.5cm]{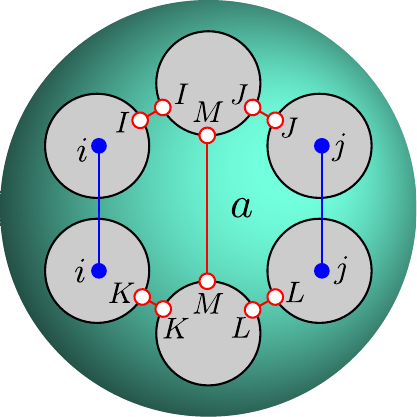}}}~.
\end{align}
The third term corresponds to the following diagram for the vertex $B_{IJK}^{(aaa)} C_{kji} D_{iI}^{(a)} D_{jJ}^{(a)} D_{kK}^{(a)}$:
\begin{align}\label{eq:V5cross}
    -\frac{2c_5}{\hbar}\frac{ g_a^{-2}\mathbb{K}_5[P_I,P_J;P_k,\bar{P}_k; P_K,P_i,P_j]}{\rho_0(k)\rho_0(\bar{k})C_0(ijk)C_0(\ib\jb\bar{k})C_0(k\bar{k}K)}
   \longrightarrow
   \vcenter{\hbox{\includegraphics[height=3.5cm]{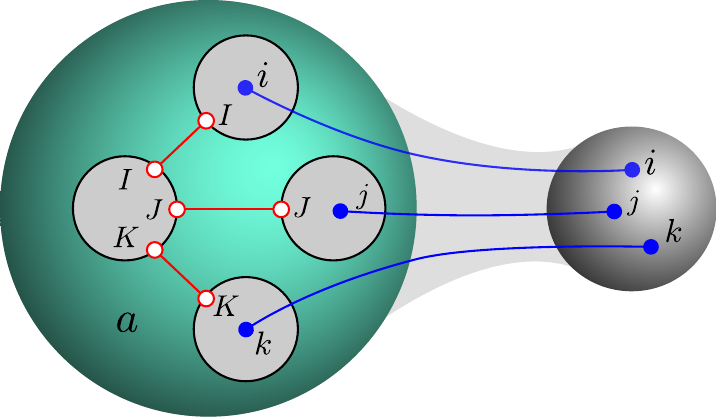}}}~.
\end{align}
The factor of two is due to the fact that the third and fourth terms are the same: Complex conjugation corresponds to mirror imaging; as the diagram corresponding to the third term is invariant under mirror imaging (along with index relabeling), it follows that the corresponding expression is real. This is an example of the utility of 3d TQFT in the sense that some 2d CFT properties become manifest in the 3d formulation. One can also check this explicitly, as commented in \cite{Runkel:1998he}.

Finally, consider $V_6$. Even though we do not know the explicit expression for $\mathbb{K}_6$, we nevertheless used the fact that it must square to one, with an inner product given by a generalization of the Verlinde inner product. To turn the block diagrams on the annulus into 3d diagrams, we essentially follow the same steps as before, except that we have two different topologies for the 3d manifold. For the bulk-channel diagram, the 3d manifold is half of the ``egg white", and for the boundary-channel diagram, the 3d manifold is a halved bagel. We then deform the skeletons just like before:
\begin{align}
    \vcenter{\hbox{\includegraphics[scale=0.6]{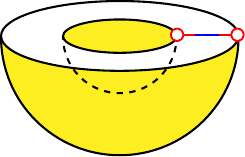}}} \longrightarrow \vcenter{\hbox{\includegraphics[scale=0.6]{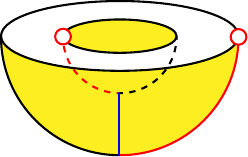}}}~,
\end{align}
and
\begin{align}
    \vcenter{\hbox{\includegraphics[scale=0.6]{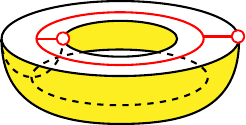}}} \longrightarrow \vcenter{\hbox{\includegraphics[scale=0.6]{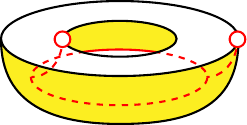}}}~.
\end{align}
The corresponding 3d diagrams are then obtained by gluing these diagrams. The first term corresponds to gluing two halved egg whites to get a whole egg white (which has two spherical boundaries), leading to the following diagram for the vertex $D_{iI}^{(a)} D_{iJ}^{(b)} D_{jI}^{(a)} D_{jJ}^{(b)}$:
\begin{align}
    \frac{c_6}{\hbar}\frac{ \delta^{2}(P_i-P_j)}{ \rho_0(i)\rho_0(\ib)C_0(i\ib I)C_0(i\ib J)}
    &\longrightarrow \vcenter{\hbox{\includegraphics[height=3.5cm]{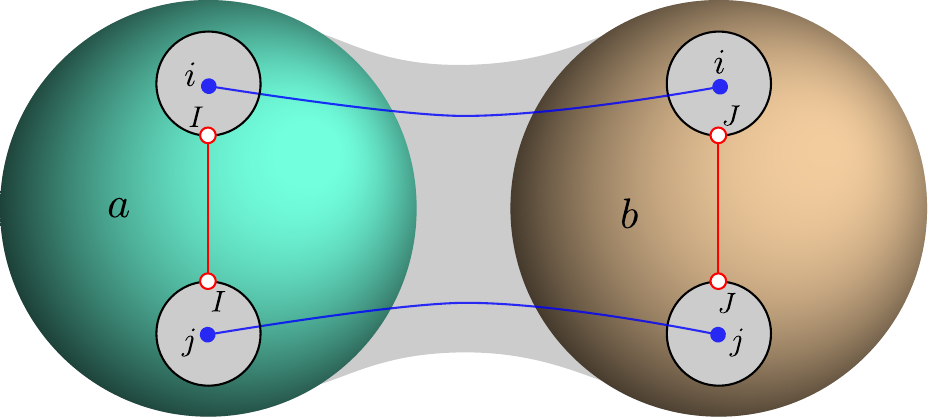}}}~.
\end{align}
The second term corresponds to gluing two halved bagels to get a whole bagel, which has a single toroidal boundary, so we have, for the vertex $B_{IKL}^{(aab)} B_{JKL}^{(bba)} B_{IMN}^{*(aab)} B_{JMN}^{*(bba)}$:
\begin{align}
    \frac{c_6}{\hbar}\frac{\delta(P_L-P_N)\delta(P_K-P_M)}{(g_a g_b)^2\rho_0(K) \rho_0(K)C_0(IKK)C_0(JKK)}
   &\longrightarrow
   \vcenter{\hbox{\includegraphics[height=3.5cm]{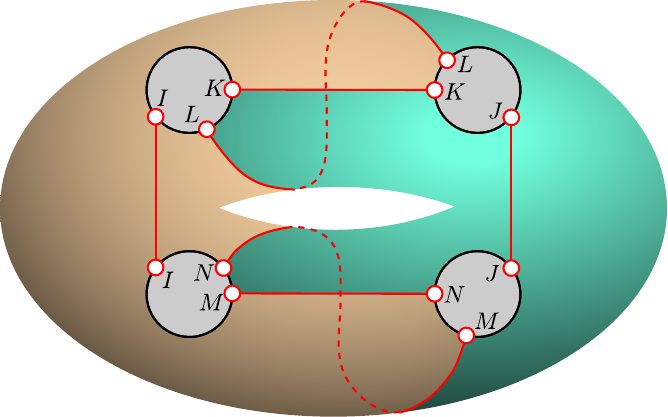}}}~.
\end{align}
The third term corresponds to gluing a halved egg white to a halved bagel, which gives a solid three-ball, i.e., for the vertex $D_{iI}^{(a)} D_{iJ}^{(b)} B_{IKL}^{(aab)} B_{JKL}^{(bba)}$:
\begin{align}
    -\frac{2c_6}{\hbar}\frac{1}{g_a g_b}  
    \frac{\mathbb{K}_{6} [P_K,P_L;P_i, \bar{P}_i;P_I, P_J] }{{\rho_0({i})\rho_0(\ib)C_0(i\ib I)C_0(i\ib J)}}
   &\longrightarrow
   \vcenter{\hbox{\includegraphics[height=3.5cm]{figs/V4cross.pdf}}}~.
\end{align}
As the third term is invariant under mirror imaging (and index relabeling), it is real, so the two cross terms are the same, hence the factor of two.  

Now, we notice that the cross term has exactly the same diagram as the cross term in $V_4$. From the TQFT point of view, this is the statement that there are two ways to cut this diagram to write it as the inner product between two conformal blocks. From this, we can read off the crossing kernel $K_6$. This fact is simple to understand in the 3d TQFT and explains the redundancy of the sixth constraint, which is not obvious from the 2d perspective. This observation is consistent with comments and results in \cite{Lewellen:1991tb,Fjelstad:2006aw}. It would be interesting to understand this redundancy more directly from the perspective of \cite{Moore:1988qv}.

With the associations above, we have completed the construction of the open-closed Virasoro TQFT. The Moore-Seiberg consistency relations \cite{Moore:1988uz,Moore:1988qv} for BCFTs ensure the theory's topological nature, i.e., the partition function on a general manifold (now with EOW branes) is independent of how we cut up the manifold into smaller pieces to compute it. It is an interesting task to verify explicitly the topological nature of the boundary sector (EOW branes), which we will do in Section~\ref{ssec:braneaction}.

Finally, let us comment on the relation to Liouville theory. The Hilbert space of the open/closed TQFT consists of the bulk Hilbert space (with both left and right sectors) and the boundary Hilbert space, each spanned by the principal series Virasoro blocks. In the closed Virasoro TQFT, Liouville CFT (as opposed to just Virasoro principal series) makes a particular appearance in its relation to the resolution of the identity, where one interprets the bra (ket) as the left (right) of the non-chiral Liouville with its spectrum of only scalar primaries. We do not have a simple analog in the open/closed case. Relatedly, the boundary condition of the open/closed TQFT is a topological one that always exists for a holomorphically factorized bulk TQFT, defined by the doubling trick. The corresponding set of boundary conditions for the ensemble of BCFTs is defined abstractly by a set of Ishibashi states. Consequently, Liouville BCFT does not make a direct appearance, and FZZT \cite{Fateev:2000ik,Teschner:2000md} or ZZ \cite{Zamolodchikov:2001ah} boundary conditions are not directly relevant. 

To understand the difference in another way, notice that one can prepare a Liouville conformal block on a compact Riemann surface $\Sigma$ with a single copy of Virasoro TQFT on $M=\Sigma \times \mathrm{Interval}$ \cite{Collier:2023fwi}, which is equivalent to having two copies of Virasoro TQFT on the $\mathbb{Z}_2$ quotient of $M$ with the reflecting boundary condition at the fixed points:
\begin{align}
    \vcenter{\hbox{\includegraphics[width=4cm]{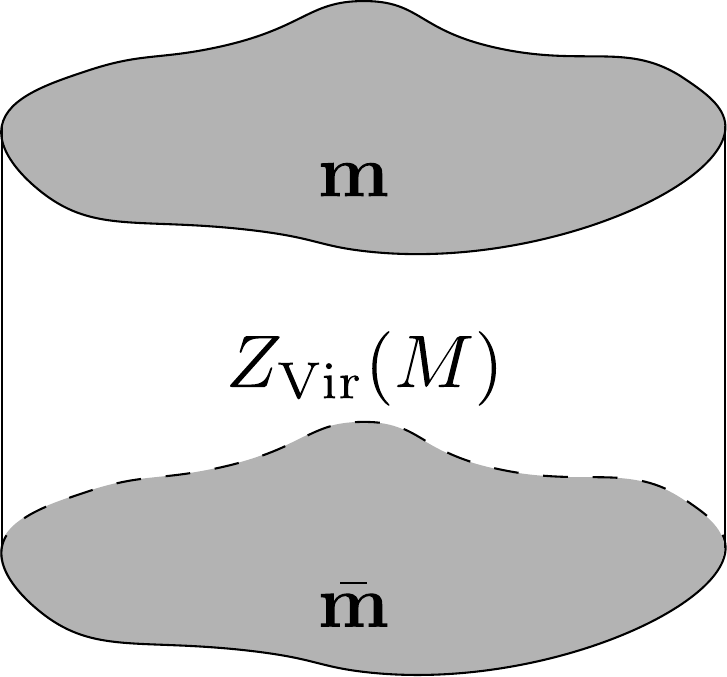}}} \quad =\quad 
    \vcenter{\hbox{\includegraphics[width=4cm]{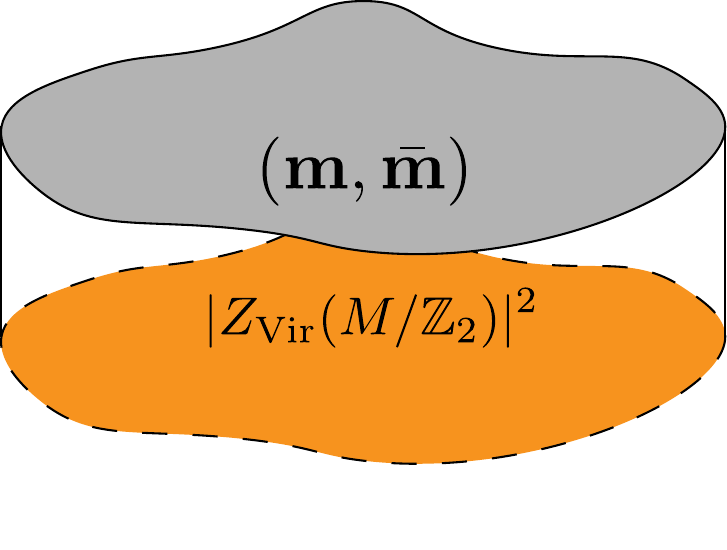}}}~.
\end{align}
Here, the gray 2d manifolds represent asymptotic boundaries, and the orange surface is the $\mathbb{Z}_2$ fixed points where we place the reflecting boundary condition. On the right-hand side, the path integral prepares a Liouville CFT state on a single asymptotic boundary $\Sigma$ with moduli $(\mathbf{m},\bar{\mathbf{m}})$. To prepare a Liouville BCFT state, take the same pictures but let $\Sigma$ be a bordered Riemann surface. The manifold before the quotient $M$ already contains codimension-one defects (EOW branes). Taking the $\mathbb{Z}_2$ quotient now introduces codimension-two objects. Formally, the choice of (local) boundary conditions here is in one-to-one correspondence with boundary conditions of the Liouville BCFT (for example, one can pick FZZT or ZZ boundary conditions), but it is non-trivial to derive them explicitly.

\subsection{\texorpdfstring{$V_0$}{V0} and surgeries}\label{ssec:sur}
There is another term in the potential of the tensor model which we have not discussed to this point: $V_0$. By construction, it specifies the leading spectrum for both the closed and the open sectors:
\begin{align}
    V_0 = V_{0,\text{closed}}+ V_{0,\text{open}}.
\end{align}
By definition, both are given by the Cardy density (with two factors of the $g$-function in the open case). As explicitly shown in \eqref{eq:Zopen-closed}, unlike $V$, $V_0$ does not have a $1/\hbar$ factor in front, so it is more leading.

In holography, the 3d manifestation of the Cardy density is straightforward. Consider CFT on a torus at high temperature. The dominant saddle in the bulk is the BTZ black hole. Geometrically, it is the statement that the dominant saddle is a solid torus where the thermal circle is contractible. 

As far as the 3d interpretation of the tensor model is concerned, the role of $V_0$ is to establish a surgery procedure. 
For $V_{0,\text{closed}}$, the corresponding surgery was explained in \cite{Jafferis:2024jkb}, which we will refer to as the \emph{toroidal surgery}. A bulk Wilson loop represents an integration over the bulk operator weights with the measure given by the Cardy density, so it is equivalent to removing a toroidal neighborhood around the loop and gluing back a solid torus where the circle parallel to the loop is contractible. In other words,
\begin{align}
    \sum_i\vcenter{\hbox{\includegraphics[height=1.5cm]{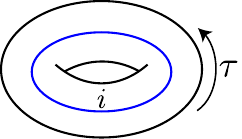}}}\equiv\int \d P\d \bar{P}|\rho_0(P)|^2\,\vcenter{\hbox{\includegraphics[height=1.5cm]{figs/surgery_closed1.pdf}}}=\vcenter{\hbox{\includegraphics[height=1.5cm]{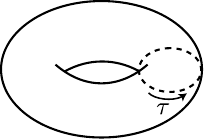}}}~.
\end{align}
Here, $\tau$ labels the circle parallel to the Wilson loop before the surgery. The toroidal surgery alters the topology of the 3-manifold and plays a crucial role in generating complicated manifolds from simple ones.

Now, consider a boundary Wilson loop with EOW branes of flavors $a$ and $b$ on either side. It represents an integral over the boundary operator spectrum with the spectral density given by
\begin{align}\label{eq:opencardy}
    g_a g_b \rho_0(P).
\end{align}
The corresponding procedure is the open analog of the toroidal surgery, which we will call the \emph{annular surgery}. Geometrically, we remove an annular neighborhood of the boundary Wilson loop from the EOW brane and glue back a geometry with the topology of the disk times an interval, i.e., a slab. In terms of pictures, 
\begin{align}
    \sum_{I\in\mathcal{H}^{ab}}\vcenter{\hbox{\includegraphics[height=1.5cm]{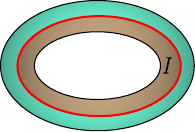}}}\equiv\int \d Pg_a g_b\rho_0(P)
    \,\vcenter{\hbox{\includegraphics[height=1.5cm]{figs/surgery_open1.pdf}}}=\vcenter{\hbox{\includegraphics[height=1.5cm]{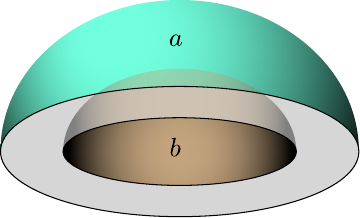}}}~.
\end{align}
Like the toroidal surgery, this changes the topology of the original manifold. In particular, it changes the topology of the EOW brane, which is also reflected in the $g$ factors, as we discuss more in Section~\ref{ssec:braneaction}. Before the surgery, an interval between a boundary with condition $a$ and another with condition $b$ is contractible to a point (on the boundary Wilson line); after surgery, this interval is no longer contractible, as can be seen from the diagram above, while the ``thermal" circle (the direction along the Wilson loop) is now contractible through the slab. This is the open analog of the BTZ black hole, and it provides the leading spectral density. 

As an example, consider a solid torus whose boundary is the EOW brane, and place a boundary Wilson loop wrapping the non-contractible cycle of the brane. The annular surgery then removes the Wilson loop along with its strip neighborhood and glues back a slab. This transforms the 3-manifold into a solid ball, and the EOW brane now has sphere topology.

Instead of summing over states running in a loop, one can also sum over states along a Wilson line connecting asymptotic boundaries. For bulk Wilson lines, we have the following surgery procedure, which we call \emph{half-toroidal}, as it can be thought of as a $\mathbb{Z}_2$ quotient of the toroidal surgery:
\begin{align}
    \sum_i\vcenter{\hbox{\includegraphics[scale=0.3]{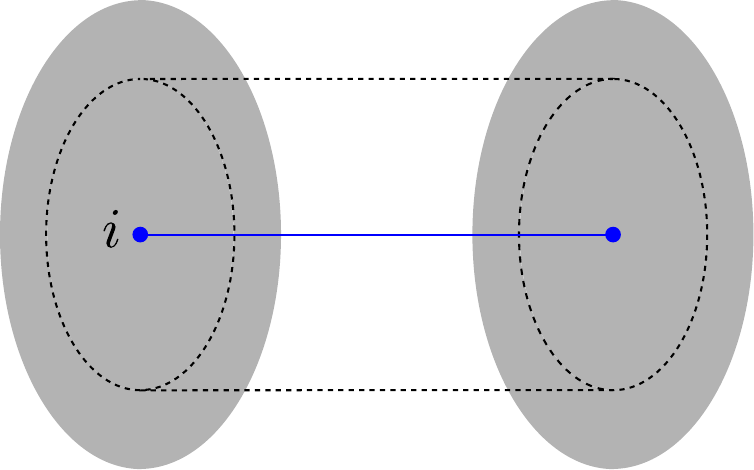}}}=\vcenter{\hbox{\includegraphics[scale=0.3]{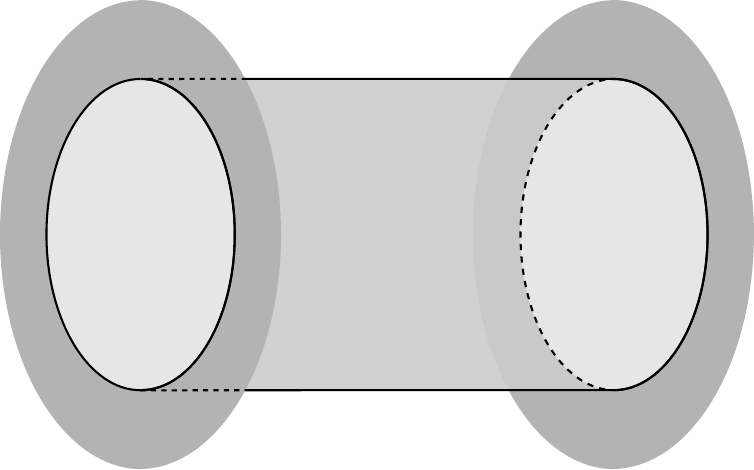}}}~.
\end{align}
On the LHS, we have a bulk Wilson line extending between two asymptotic regions, and the dashed lines outline a cylinder neighborhood of the Wilson line. To arrive at the RHS, we simply remove this cylinder. The disk neighborhood of each endpoint of the Wilson line is now gone, and the remaining boundary of the cylinder becomes asymptotic. In other words, the sum turns the Wilson line into a tube whose annular boundary is now part of the asymptotic boundary. The two asymptotic boundaries become connected; if they are already connected, the genus of the asymptotic boundary increases by one.

For boundary Wilson lines extending between asymptotic boundaries, we have the open analog of the above, which we call \emph{half-annular}, as it is a $\mathbb{Z}_2$ quotient of the annular surgery. Pictorially,
\begin{align}
    \sum_{I\in\mathcal{H}^{ab}}\vcenter{\hbox{\includegraphics[scale=0.35]{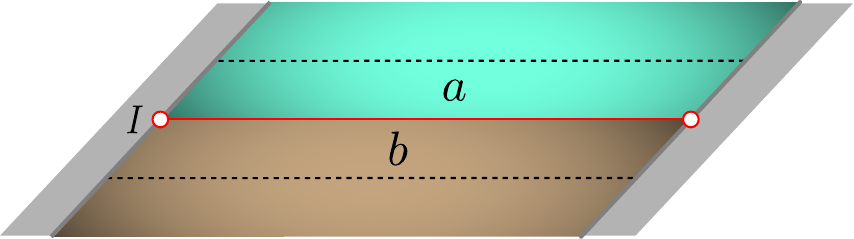}}}=\vcenter{\hbox{\includegraphics[scale=0.35]{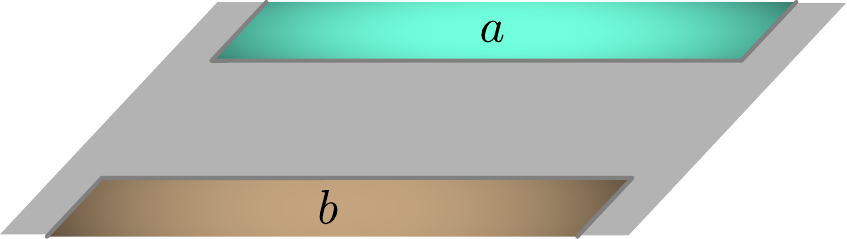}}}~,
\end{align}
where the gray regions depict asymptotic boundaries. The surgery joins the two asymptotic boundaries by turning the strip neighborhood of the boundary Wilson line into a part of the asymptotic boundary; if they are already connected, the surgery increases the number of circular borders by one, e.g., a disk asymptotic boundary turns into an annulus. 

In the purely closed case \cite{Jafferis:2024jkb}, all manifolds are produced by the tensor-matrix model. In the open-closed case, we can now ask a similar question: Are all manifolds with EOW branes produced? More precisely, given prescribed asymptotic boundaries, does the model produce all possible bordisms with these boundaries? We expect the answer to be affirmative, and it would be interesting to show it, perhaps using the quadrupling trick explained in \cite{Wang:2025bcx}.

\subsection{Off-shell geometries}\label{ssec:offshell}
The conformal blocks associated with the empty torus are not normalizable with respect to the Verlinde inner product. This is related to the fact that the corresponding 3d manifolds (e.g., the torus wormhole) have divergent partition functions when evaluated using Virasoro TQFT. However, these manifolds do have finite 3d gravity partition functions because one needs to quotient the TQFT answers by the mapping class group. 

In the closed case, the matrix part of the model makes predictions for off-shell geometries in pure 3d gravity. In particular, it gives the correct answer for the torus wormhole \cite{Cotler:2020ugk}. We have constructed the matrix part of the BCFT model analogously, and we are going to see that it makes predictions for off-shell 3d manifolds with branes. 

Let us recall the annulus potential \eqref{eq:Vann_long}:
\begin{align}
    V_{\rm annulus}&=\sum_{a,b}\sum_{I,J} {\tilde{K}(P_I,P_J)} 
    - 2\sum_{a,b}\sum_{I} g_a g_b \int {\d P \tilde{K}}(P,P_I) \mathbb{S}_{\mathbbm{1} P}[\mathbbm{1}]
    \nn
    &- 2\sum_{a,b}\sum_{i}'\sum_{I} A_{i}^{(a)} A_{i}^{(b)} \int {\d P \tilde{K}}(P,P_I) \mathbb{S}_{P_i P}[\mathbbm{1}]\nn
    &+2\sum_{a,b}\sum_{i}' A_{i}^{(a)} A_{i}^{(b)} g_a g_b \int {\d P \d P'}  \mathbb{S}_{P_i P}[\mathbbm{1}] {\tilde{K}}(P,P') \mathbb{S}_{P' \mathbbm{1}}[\mathbbm{1}]
    \nn
    &+\sum_{a,b}\sum_{i,j}' A_{i}^{(a)} A_{i}^{(b)} A_{j}^{(a)} A_{j}^{(b)} \int {\d P \d P'}  \mathbb{S}_{P_i P}[\mathbbm{1}] {\tilde{K}}(P,P') \mathbb{S}_{P' P_j}[\mathbbm{1}],
\end{align}
There are five terms. The first one is the kinetic term, so we take its functional inverse to obtain the propagator. It is then assigned the following geometry which we call the annulus wormhole: 
\begin{align}
    \langle \rho^{ab}(P_I)\rho^{ab}(P_J)\rangle\equiv {\tilde{K}^{-1}(P_I,P_J)} \longrightarrow \vcenter{\hbox{\includegraphics[scale=0.4]{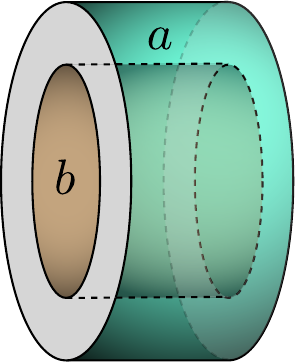}}}~.
\end{align}
This is the open analog of the torus wormhole, and it would be interesting to reproduce this answer directly from a bulk computation along the lines of \cite{Cotler:2020ugk}. Notice that $I$ and $J$ both live in the open Hilbert space $\mathcal{H}^{ab}$ where $a$ and $b$ are generally different. This is reflected in the diagram by the fact that the two branes can have different flavor labels. Notice that when $a=b$, there is a single term in the sum, while there are two when $a\ne b$. The 3d perspective is consistent with this observation: There is only one way of connecting the two annuli if $a\ne b$ but two otherwise. 

The second term is single trace, and the 3d diagram associated to it is a simple slab:
\begin{align}
    g_a g_b \int {\d P \tilde{K}}(P,P_I) \mathbb{S}_{\mathbbm{1} P}[\mathbbm{1}]\longrightarrow\vcenter{\hbox{\includegraphics[scale=0.4]{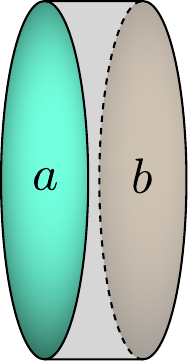}}}~.
\end{align}

The third term is interesting because it does not have an analog in the closed case. It is an interaction term between the open spectrum and the dynamical data $A_i$. The corresponding 3d diagram looks like
\begin{align}
    A_{i}^{(a)} A_{i}^{(b)} g_a g_b \int {\d P \d P'}  \mathbb{S}_{P_i P}[\mathbbm{1}] {\tilde{K}}(P,P') \mathbb{S}_{P' \mathbbm{1}}[\mathbbm{1}]\longrightarrow\vcenter{\hbox{\includegraphics[scale=0.4]{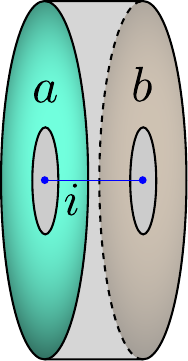}}}~.
\end{align}

The remaining two terms are purely interaction terms for $A$. For the fifth one, we have the diagram
\begin{align}
    A_{i}^{(a)} A_{i}^{(b)} A_{j}^{(a)} A_{j}^{(b)} \int {\d P \d P'}  \mathbb{S}_{P_i P}[\mathbbm{1}] {\tilde{K}}(P,P') \mathbb{S}_{P' P_j}[\mathbbm{1}]
    \longrightarrow\vcenter{\hbox{\includegraphics[height=2cm]{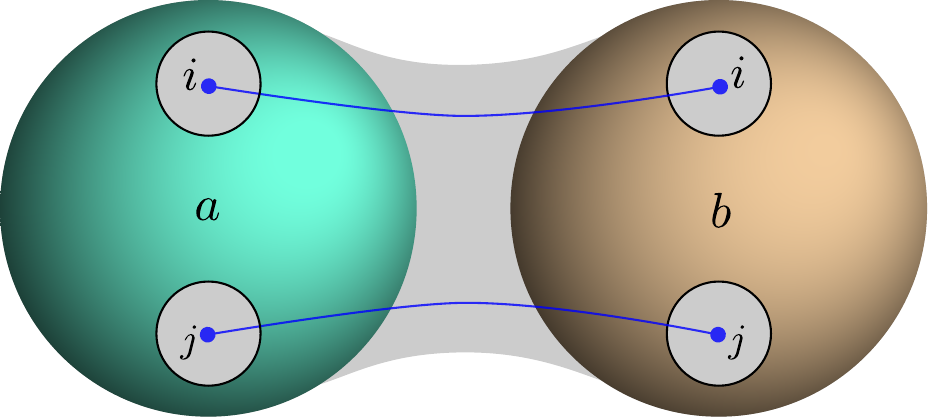}}}~,
\end{align}
and for the fourth one, the diagram is obtained by removing one of the Wilson lines in the diagram above (along with the disks at its ends).

Compared to the closed matrix model, the 3d diagrams associated to the open matrix model are simpler, as it is not graded by spin. This means that we do not have something like the Dehn twist to perform. Owing to the Vandermonde, the matrix model systematically computes all contributions to the density correlator, which are well-known to be associated to 2d surfaces. In our model, the corresponding 3d manifolds are simply those 2d diagrams times an interval. For example, the matrix model produces a connected contribution to $\langle\rho^{ab}(P_I)\rho^{ab}(P_J)\rho^{ab}(P_K)\rangle$ that geometrically corresponds to a ``thickened'' pair of pants, i.e., a pair of pants times an interval:
\begin{align}
    \vcenter{\hbox{\includegraphics[height=3cm]{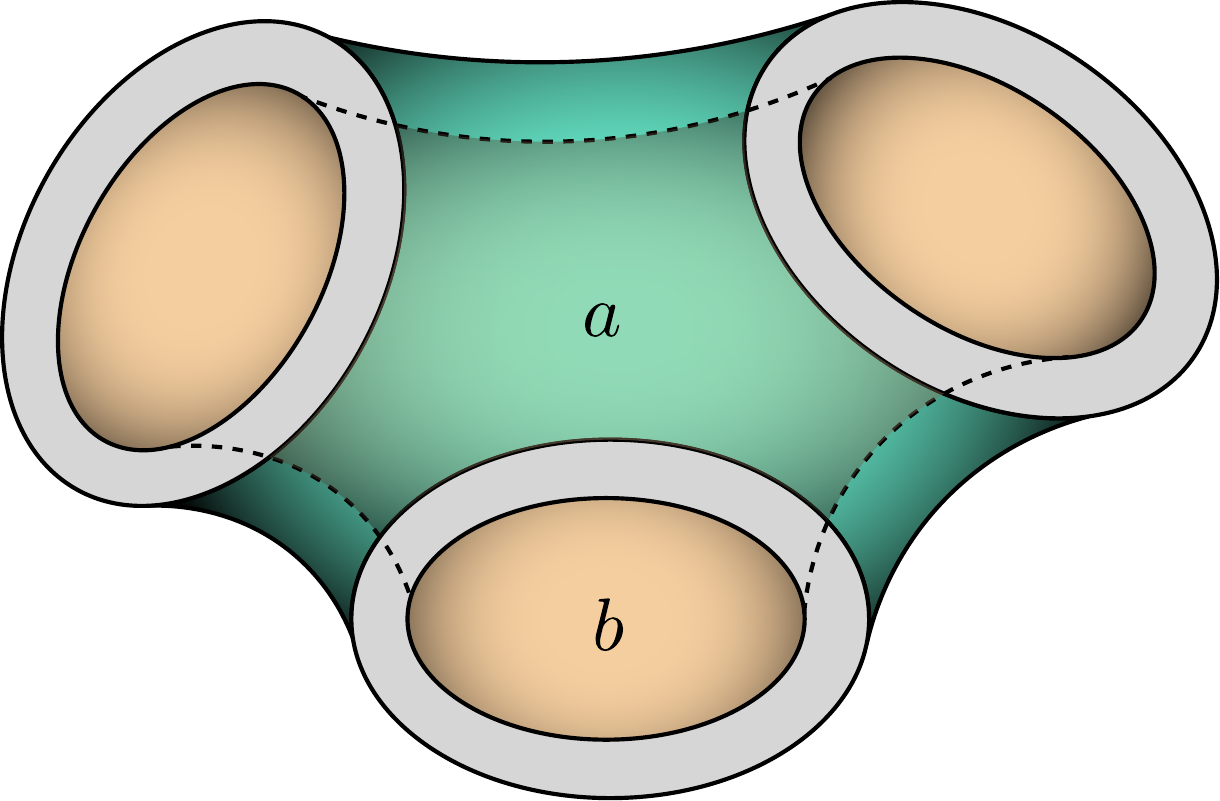}}}~.
\end{align}
Consequently, whenever there are three boundary Wilson loops in between branes with colors $a$ and $b$, removing their annular neighborhoods (trenches) and gluing the thickened pair of pants produces a subleading contribution. 

If we set all the $g$ functions associated to the EOW branes to one, i.e., removing the dependence on the brane topology and their color labels, we obtain Virasoro Minimal String (VMS) \cite{Collier:2023cyw}, as can be seen from the spectral density \eqref{eq:opencardy}. It is explained in \cite{Collier:2023cyw} that VMS is related to chiral 3d gravity on $\Sigma_{g,n}\times S^1$. See also \cite{Maloney:2015ina,Eberhardt:2022wlc}. In contrast, our bulk description is non-chiral 3d gravity on manifolds with topology $\Sigma_{g,n}\times \mathrm{Interval}$. It would be interesting to establish a connection between the two.

\subsection{Brane action}\label{ssec:braneaction}
The closed Virasoro TQFT provides a precise quantization of 3d pure gravity. Since we have now constructed, in a bottom-up way, an open-closed extension of Virasoro TQFT, we can ask the question: What classical theory is this TQFT a quantization of? 

The classical theory should now have a codimension-one object whose quantization gives the EOW branes in the TQFT. The nontrivial task is to figure out the action, if any. 

We now claim that the classical (Euclidean) action is given by (along with the standard GHY term and counter terms which we are omitting)
\begin{align}\label{eq:action_EOW}
    I=-\frac{1}{16\pi G_N} \int_{\mathcal{M}} \sqrt{g} \,(R - 2\Lambda)- \frac{1}{8\pi G_N} \sum_a\int_{Q_a} \sqrt{h}\, K- \sum_a\chi_a \log g_a,
\end{align}
where $\mathcal{M}$ is a 3d manifold, $Q_a$ is a 2d surface which we call an EOW brane, $a$ labels different EOW branes, $g_a$ is a constant associated to each EOW brane, and $\chi_a$ is the Euler characteristic of $Q_a$. In the absence of asymptotic boundaries, $\partial \mathcal{M} = \sqcup_a Q_a$, and each $Q_a$ is a closed surface. With asymptotic boundaries, say $\mathcal{A}_i$, we have $\partial\mathcal{M} = (\sqcup_i\mathcal{A}_i)\sqcup(\sqcup_a Q_a)$. Now, each $\mathcal{A}_i$ can be either closed (corresponding to CFTs on compact Riemann surfaces) or open (corresponding to CFTs on bordered Riemann surfaces), and each $Q_a$ can be either closed or end on one or more $\partial\mathcal{A}_i$'s. 

At the EOW branes, Neumann boundary conditions should be imposed. The Neumann boundary condition sets $K=0$ at the brane. With $K=0$, we can glue two 3d on-shell manifolds along the brane to produce another on-shell manifold. For example, gluing two open 6$j$ manifolds gives the closed 6$j$ manifold. The assigned partition functions are consistent with this fact, up to factors of $g_a$ which we discuss next. 

To see why we have a topological term for the branes, it is easiest to look at the partition function on some manifold without asymptotic boundaries. This can be done via an exhaustive approach. As an example, consider the diagram \eqref{eq:V5cross}. Gluing two of them using the propagators would lead to a manifold whose boundary is a single genus-five brane along with some bulk and boundary Wilson lines. Next, perform surgeries for the Wilson lines. Toroidal surgery on the bulk Wilson lines changes the topology of the 3d manifold but does not change the topology of the brane. Annular surgeries change the topology of the brane. Since there are five boundary Wilson loops, we do it five times which removes all five genus holes. The resulting surface is a sphere, which has Euler characteristic $\chi_a=2$. Looking at the $g_a$ dependence in the functions, we obtain a factor of $1/g_a^4$ from each of the two vertices and a factor of $g_a^2$ from each annular surgery. Together, the $g$ dependence is given by
\begin{align}
    ({g_a^{-4}})^2 (g_a^2)^5 = g_a^2=g_a^{\chi_a},
\end{align}
as predicted by the action \eqref{eq:action_EOW}. 

We now compare and contrast this topological action with the model of \cite{Wang:2025bcx}. The action \eqref{eq:action_EOW} is technically a special case of the action in \cite{Wang:2025bcx}. For example, there the EOW branes with different labels can join at sharp corners (kinks), and the contribution of each kink is proportional to the 1d Euler characteristic of the kink with a coefficient that depends on the boundary conditions on both sides of the kink. Our model does not involve kinks because the boundary Wilson lines all carry weights above the threshold. However, we can analytically continue the weights below the threshold and easily verify that the TQFT partition functions we derived earlier have the exact $g$ factors computed by the model of \cite{Wang:2025bcx}. A difference between \eqref{eq:action_EOW} and \cite{Wang:2025bcx}, however, is that the latter only worked with on-shell manifolds. For on-shell manifolds, the topological action agrees with the one proposed in \cite{Takayanagi:2011zk,Fujita:2011fp} as the bulk dual for a single BCFT, as shown in \cite{Geng22,Wang:2025bcx}. In other words, \cite{Wang:2025bcx} extends the formalism of AdS/BCFT to allow multiple boundaries at large $c$, and we extend it further to include both on-shell and off-shell manifolds by working with finite $c$.

\section{Purely open bootstrap}\label{sec:pureopen}

As explained in the introduction, one interesting application of the tensor model is to a purely open bootstrap problem. Consider for example a CFT partition function on a closed Riemann surface and perform a triangulation. We can then remove holes at the vertices to turn it into a BCFT partition function (only a subregion is drawn):
\begin{align}\label{eq:trig}
    \vcenter{\hbox{\includegraphics[height=2.5cm]{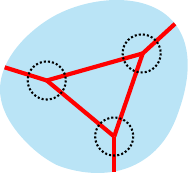}}}
    \quad\longrightarrow\quad
    \vcenter{\hbox{\includegraphics[height=2.5cm]{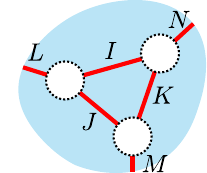}}}
    \quad\longrightarrow\quad
    \vcenter{\hbox{\includegraphics[height=2.5cm]{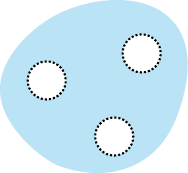}}}~.
\end{align}
In the second diagram, the red intervals are places where a complete basis of boundary operators (open states) is inserted. Each region enclosed by three red intervals and the boundaries of the holes corresponds to a boundary-to-boundary OPE coefficient $B_{IJK}^{(abc)}$, where $a$, $b$, and $c$ label the boundary conditions imposed at the holes. We can pick all of them to be the same so that we can drop these labels.  In the third diagram, the red intervals are removed, suggesting that we do not need to pick a channel to specify the BCFT partition function, as a consequence of four-point crossing on the disk. We can evaluate this in the following model.

Taking the open-closed tensor model \eqref{eq:Zopen-closed} but keeping only the boundary data, we obtain the following purely open tensor model:
\begin{equation}\label{eq:Zopen}
\mathcal{Z}= \int 
    \mathcal{D} [h_I,B_{IJK}] \,
    \e^{-V_0(h_I)-\frac{1}{\hbar} V_3(h_I,B_{IJK})},
\end{equation}
where $V_3$, given in \eqref{eq:V3expanded}, is the potential from the disk four-point crossing constraint. All other terms in the potential, including $V_{\rm annulus}$, involve bulk (closed) data, so they are absent in the purely open model.

We can then proceed to look for 3d manifolds whose asymptotic boundary is given by the BCFT surface, i.e., the third diagram in \eqref{eq:trig}. This resonates with the ideas in \cite{VanRaamsdonk:2018zws,Chen:2024unp,Hung:2024gma,Bao:2024ixc}. These 3d manifolds will necessarily contain EOW branes. There will be EOW branes that end on the holes, but there can also be ones that do not. Focusing on the topology of the EOW branes, some examples are given by
\begin{align}\label{eq:holes}
    &\vcenter{\hbox{\includegraphics[height=1cm]{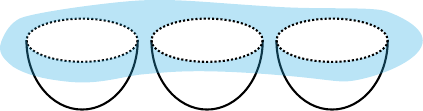}}}
    +
    \vcenter{\hbox{\includegraphics[height=1cm]{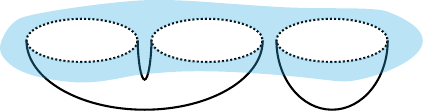}}}+
    \vcenter{\hbox{\includegraphics[height=1cm]{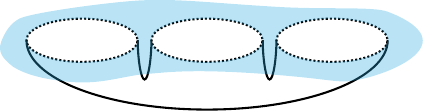}}}
    \\
    +\,&
    \vcenter{\hbox{\includegraphics[height=1cm]{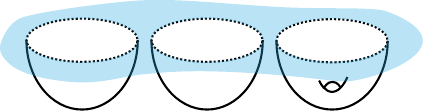}}}
    +
    \vcenter{\hbox{\includegraphics[height=1cm]{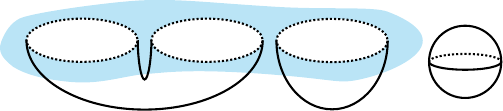}}}+
   \vcenter{\hbox{\includegraphics[height=1cm]{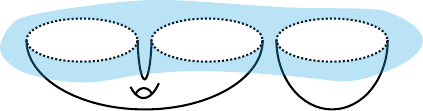}}}+\cdots.\nonumber
\end{align}
These are connected 3-manifolds with potentially more than one boundary component. Only the topologies of the EOW branes (white) are shown, so for each diagram above, there are many 3d manifolds with different 3d topologies.

There is a beautiful connection between the 3-manifolds constructed from the open $6j$ diagrams of the purely open tensor model and the idea of tetrahedral decomposition. It follows from the doubling trick applied to surgeries on the chain-mail link associated to a tetrahedral decomposition \cite{Roberts1995}, as described in the context of tensor models in \cite{Jafferis:2024jkb}. Here, the doubling trick means gluing a given 3d manifold with EOW branes to its mirror image along the brane. Mirror imaging also complex conjugates the partition function. 

More precisely, given an open tensor model diagram with only $6j$ vertices, one can construct a tetrahedral decomposition as follows. Each vertex is associated to a truncated tetrahedron (corners removed). Propagators in the tensor model map to the gluing of tetrahedra along their faces. Note that in the open case, only cyclic permutations are permitted, corresponding to the necessity of gluing tetrahedra on opposite sides of a face, a desirable feature. (Gluing two tetrahedra on the same side of a face would create a non-smooth surface in the 3d manifold.) External tensor insertions remain unglued asymptotic boundaries. 

The vertices of the tetrahedral gluing will be linked by 2-manifolds that need not be spheres (unlike the following example). Since the tetrahedra are truncated, these become EOW brane boundaries. The resulting manifold is identical to the one obtained by gluing open $6j$ manifolds and performing the annular surgery on all Wilson loops. (This follows from doubling along the EOW brane boundaries, which results in the chain-mail link construction built out of closed $6j$ manifolds. That produces $M \#_{V} \bar{M}$, the connected sum of $M$ with its mirror image over all the links of the vertices in the triangulation. Its $\mathbb{Z}_2$ quotient is the above.)

As an example, consider the 4--1 Pachner move, the procedure of gluing four (truncated) tetrahedra along their hexagonal faces as shown:
\begin{align}
    \vcenter{\hbox{\includegraphics[height=4cm]{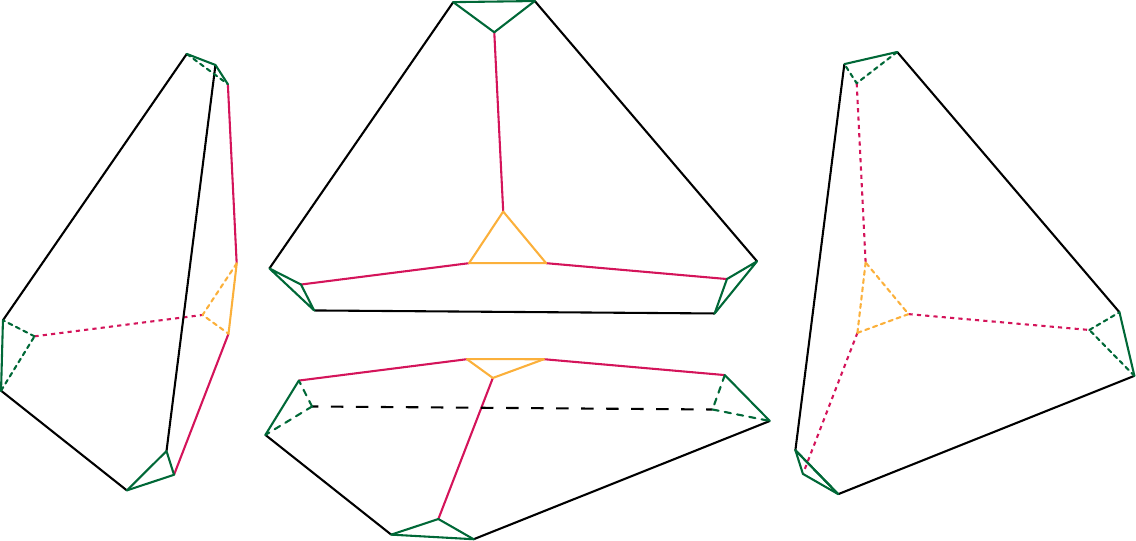}}}\longrightarrow
    \vcenter{\hbox{\includegraphics[height=3.5cm]{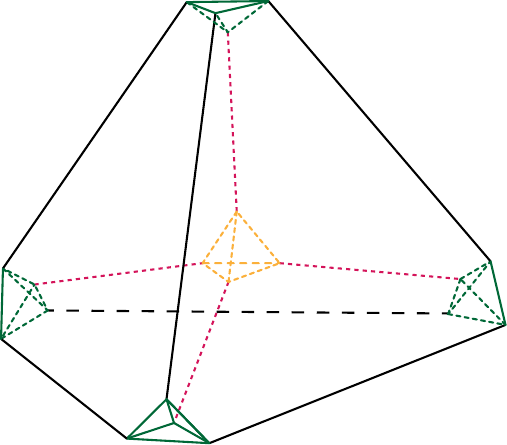}}}~,
\end{align}
where the yellow triangular surfaces join to form a boundary of spherical topology in the resulting manifold, and the green triangles form disks. The resulting manifold is another truncated tetrahedron but with an interior boundary. In terms of the open $6j$ diagrams, this corresponds to joining four open $6j$ manifolds using six $B$-propagators, each propagator representing a face gluing:
\begin{align}
    \vcenter{\hbox{\includegraphics[height=4.5cm]{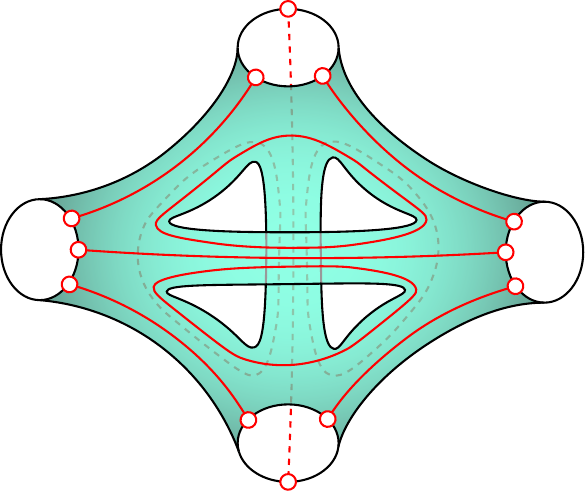}}}\longrightarrow
    \vcenter{\hbox{\includegraphics[height=4.5cm]{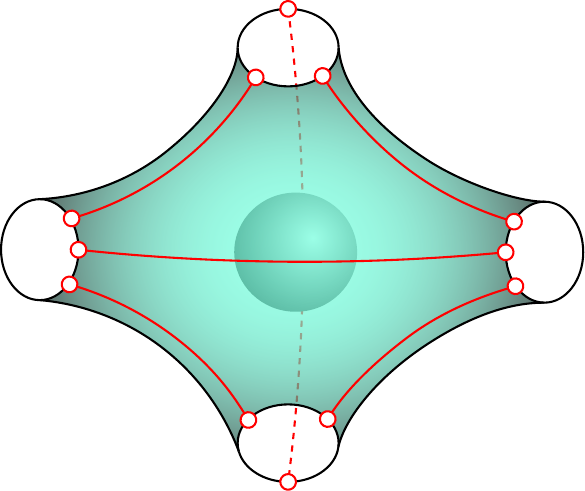}}}~,    
\end{align}
where we have performed the annular surgery for each of the resulting boundary Wilson loops (two shown on the front and two at the back of the diagram). With a bit of mental exercise, one sees that the surgery turns it into another open $6j$ manifold but with an interior spherical EOW brane, consistent with the result of the 4--1 Pachner move. 

Since each EOW brane contributes a factor of $g^\chi$, for large $g$, spherical branes dominate. However, the contributions from spherical branes (such as the example above) as evaluated by the tensor model gluing rules are divergent. If these can be computed and subtracted off, i.e., renormalized away, then the dominant contributions come from the first diagram of \eqref{eq:trig} (where all branes are disks that end on the holes). We then get a factor of $g^V$ where $V$ is the number of holes, or the number of vertices in the triangulation. The associated tetrahedral decomposition would have all vertices at the boundary.  From the tensor model perspective, as the fundamental observables are the OPE coefficients, we will be evaluating the middle diagram of \eqref{eq:trig}, i.e., in a particular channel decomposition. This means that we would require a slightly stronger statement than the existence of a triangulation without internal vertices: For a given triangulation of the boundary manifold, we would like to ask if there is such a triangulation of the 3d manifold whose boundary coincides with the boundary triangulation. It would be interesting to answer this question.

It is therefore useful to understand the divergent nature of the contributions from the spherical EOW branes. As explained in the introduction, these are analogous to divergences in 3d manifolds without EOW branes that have an $S^2$ handle. One expects that understanding these divergences might also shed light on those divergences. However, this is a difficult question that is beyond the scope of the paper. We will only give an example below.

Start with \eqref{eq:V3diag}, connecting $B_{KLM}$ and $B_{NLK}$ with a $B$-propagator. We then get a torus brane with two boundary Wilson loops along the non-contractible cycle. Doing the annular surgery described in Section~\ref{ssec:sur} on one of these loops removes the Wilson loop and turns the EOW brane into a sphere. Doing the annular surgery for the other Wilson loop then turns the manifold into \eqref{eq:propB} with a 3-ball removed from the interior. Using the TQFT rules, the resulting partition function is then
\begin{align}
    \frac{1}{C_0(IJM)}\delta(0)\left[\int \d P \rho_0(P)\right]^2.
\end{align}
Relative to \eqref{eq:propB} (not as a propagator but as a manifold), we have an overall divergent factor that is independent of the rest of the manifold. 

If we can understand the divergences, we can then try to answer the important question of whether the topological expansion reproduces the original CFT partition function computed using the purely closed tensor model. In particular, for finite $g$, we do not get parametric suppression of higher EOW brane topologies, but 3d manifolds with higher EOW brane topologies typically have a greater volume, which are therefore suppressed in the $\e^{-c}$ expansion. 

One may worry that an additional constraint might be needed to enforce that all holes close properly in the shrinking limit. In other words, we want to ensure that the partition function scales with a factor of $\e^{\frac{\pi c}{6\epsilon}}$ in the limit the size of the hole $\epsilon$ goes to zero for each hole \cite{Chen:2024unp}. We argue that this is already taken care of by $V_0$, the leading spectral density appearing in the tensor potential. To see this, consider an annulus with one boundary operator. This is a special case of \eqref{eq:cstr6block} with $J$ set to $\id$. Some details can be found in e.g.~\cite{Numasawa:2022cni}. In this limit where the circle without insertions shrinks, the closed channel becomes a disk one-point function, which vanishes unless $I=\id$, in which case the equation becomes the empty torus constraint \eqref{eq:cstr_ann}. This gives the leading density of the boundary spectrum, which is part of $V_0$ by construction. More general surfaces with holes can be reduced to this situation via crossing symmetry of the four-point disk function. For example, consider the disk two-point function with a hole removed. Crossing turns it into an open pair of pants glued to a disk one-point function with a hole:
\begin{align}
    \vcenter{\hbox{\includegraphics[height=2cm]{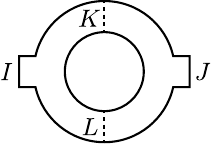}}}
    \quad\longrightarrow\quad
    \vcenter{\hbox{\includegraphics[height=2.5cm]{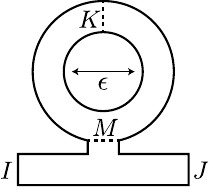}}}~.
\end{align}
Since the disk one-point function with a hole shrinks properly as argued earlier, this example works automatically.

\section{Non-orientable CFTs}\label{sec:xcft}

Up to this point, we have restricted ourselves to bootstrap on orientable Riemann surfaces. In this section, we discuss how to generalize the model even further to include non-orientable CFTs (on both compact and bordered Riemann surfaces). We will list all the ingredients for the construction and demonstrate it explicitly for one of the new terms in the potential.

First of all, let us discuss the new variables. In addition to all the BCFT data reviewed in Section~\ref{ssec:bdata}, the main extra piece of data needed for non-orientability is the one-point function of a bulk operator on the crosscap ($\mathbb{RP}^2$). Let $X_i$ denote this data for a scalar bulk operator with conformal weights $(h_i,\bar{h}_i)$. Pictorially, 
\begin{align}\label{eq:dataX}
   X_{i} \longrightarrow \vcenter{\hbox{\includegraphics[height=2cm]{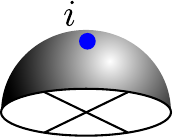}}}~.
\end{align}

Parity plays an interesting role in the non-orientable case. First of all, parity means that operators with opposite spins are paired up. We can therefore restrict to positive spins. Secondly, for scalars, we distinguish between parity-odd and parity-even states. We will use the notation $K_i=\pm1$ for the eigenvalues of parity.\footnote{With only Virasoro symmetry, $K_i$ equals the eigenvalue of involution, so we do not distinguish them \cite{Pradisi:1995pp,Stanev:2001na,Tsiares:2020ewp}.} Thirdly, $X_i$ is only non-zero for even-parity scalar operators ($h_i=\bar{h}_i$ and $K_i=1$). Finally, $C_{ijk}$ is only non-zero if an even number of the three states are parity-odd (none or two).

With the variables clarified, let us now discuss bootstrap. There are three additional bootstrap constraints, making it a total of nine \cite{Fioravanti:1993hf}. These come from crossing symmetries of the following three objects: two bulk operators on the crosscap, one bulk operator on the Klein bottle, and one bulk operator and one boundary operator on the M\"obius strip. The first one and a special case of the second one have been studied recently in the context of CFT universal asymptotics in \cite{Tsiares:2020ewp}. 

Let us study the crosscap constraint as an example. In terms of conformal blocks, the crossing symmetry states \cite{Fioravanti:1993hf,Pradisi:1995pp}
\begin{align}
    \sum_s C_{ijs} X_s \mathcal{F} \begin{bmatrix}
        P_j & P_i \\ \bar P_i & \bar P_j
    \end{bmatrix} (P_s | \eta) = K_i (-1)^{h_i-\bar h_i + h_j - \bar h_j} \sum_t C_{ijt} X_t \mathcal{F} \begin{bmatrix}
        P_j & \bar P_i \\  P_i & \bar P_j
    \end{bmatrix} (P_t | 1-\eta),
\end{align}
where $\eta$ is a real cross ratio. As before, we can project this on a particular $s$-channel block labeled by $P$ to express this equation as
\begin{align}
    \mathcal{M}_{7,ij}(P)=\sum_{m}\left(K_i (-1)^{s_i + s_j}  C_{ijm} X_m \delta(P - P_m) - C_{ijm} X_m \mathbb{F}_{m P} \begin{bmatrix}
         j & i \\\ib & \jb
     \end{bmatrix} \right).
\end{align}
Notice that, since $X_m$ is only non-vanishing for scalar operators $\phi_m$, the sum over $m$ reduces to a sum over scalars. 

To construct the tensor model, we modify it to
\begin{align}
    \mathcal{Z}=\prod_{s \in \mathbb{Z}} \int 
    \mathcal{D} [\Delta_{s},h_I,C_{i j k},B_{IJK},D_{iI} ,A_{i},X_i] \,
    \e^{-V_0(\Delta_s,h_I)-\frac{1}{\hbar} V(\Delta_s, h_I,C_{i j k},B_{IJK},D_{iI},A_{i},X_i)},
\end{align}
where 
\begin{align}
    V = \sum_{i=1}^9 V_i + V_{\rm torus} + V_{\rm annulus}.
\end{align}
The three new terms $V_7$, $V_8$, and $V_9$ come from the three additional constraints mentioned earlier. We have suppressed the boundary condition labels.

Squaring the seventh constraint gives a new term in the tensor potential:
\begin{equation}
     V_7=c_7
     \sum_{i,j}
    \int_{\mathbb{R}_+ \cup \mathds{1}}\mathcal{M}_{7,ij}(P)^*\mathcal{M}_{7,ij}(P)\, \mu_7(P) {\d P},
\end{equation}
where
\begin{equation}
    \mu_7(P)=\frac{1}{\rho_0({P})C_0(ijP)C_0(\ib\jb P)}.
\end{equation}

Expanding $V_7$ gives
\begin{align}\label{eq:V7expanded}
V_7= 2c_7\sum_{i,j,m,n}
& C_{ijm}  C_{nji} X_{m}X_{n}
\\
&
\Bigg(\frac{\delta(P_m-P_n)}{\rho_0(m) C_0(ijm) C_0(\ib\jb m)} 
-\frac{K_i (-1)^{s_i + s_j}}{C_0(ijm)C_0(\ib\jb m)  C_0(\ib jn)C_0(i\jb n) }\left\{\begin{array}{lll}
n & \jb & i \\
m & j & \ib
\end{array}\right\}\Bigg).\nonumber
\end{align}
The first term corresponds to a 3d manifold that connects two three-punctured spheres and two one-punctured crosscaps:
\begin{align}\label{eq:V7diag}
   \frac{C_{ijm}  C_{nji} X_{m}X_{n}\delta(P_m-P_n)}{\rho_0(m) C_0(ijm) C_0(\ib\jb m)} 
    &\longrightarrow \vcenter{\hbox{\includegraphics[height=4cm]{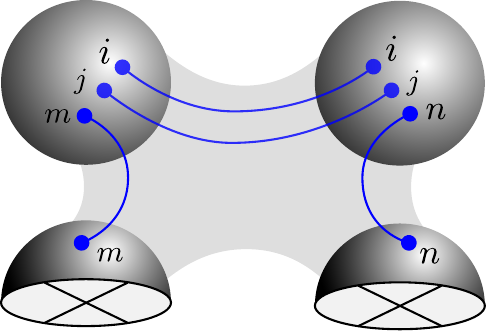}}}~.
\end{align}
The picture is schematic, as it does not specify the topology of the interior. Compared to the orientable case, the construction of such diagrams is less systematic and more difficult to visualize. Nevertheless, it is possible to make progress by thinking about these objects in the covering space. As a result, we propose that this manifold has the topology of $\mathbb{RP}^2$ times an interval with two balls removed. This is a non-orientable manifold. Similarly, the second term corresponds to, again schematically:
\begin{align}\label{eq:V7cross}
   \frac{C_{ijm}  C_{nji} X_{m}X_{n}K_i (-1)^{s_i + s_j}}{C_0(ijm)C_0(\ib\jb m)  C_0(\ib jn)C_0(i\jb n) }
    \left\{\begin{array}{lll}
    n & \jb & i \\
    m & j & \ib
    \end{array}\right\} 
    &\longrightarrow \vcenter{\hbox{\includegraphics[height=4cm]{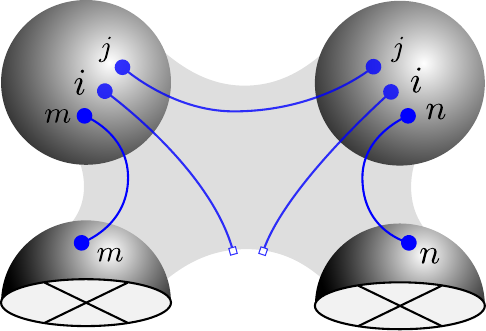}}}~.
\end{align}
To understand its topology, it is easier to remove all Wilson lines except the one labeled by $i$ and shrink the spheres to points. Then the manifold has the topology of $\mathbb{RP}^2$ times an interval, plus a Wilson line. Now deform the Wilson line so that it lies entirely at the mid-point of the interval. For convenience, also put the two points at the same location. The cross-section at the mid-point of the interval now looks like
\begin{align}
   \vcenter{\hbox{\includegraphics[height=2cm]{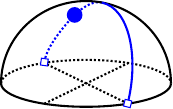}}}~.
\end{align}
This Wilson loop is non-contractible on $\mathbb{RP}^2$. In contrast, if we repeat this for $j$, we get a contractible loop.

We can construct the remaining terms in the potential $V_8$ and $V_9$ similarly from the remaining two bootstrap constraints. However, as the corresponding crossing kernels have not been explicitly worked out, we will not achieve it in this work. The corresponding 3d diagrams will also be more complicated. In particular, $V_9$ would be associated to non-orientable 3d manifolds with EOW branes. It is not clear to us whether the EOW branes themselves can be non-orientable 2d surfaces.

From the second term, we can isolate the kinetic term for $X_i$. Setting $i=\id$ or $j=\id$, we obtain
\begin{align}
   K_X = -4c_7\sum_{i}'\frac{  X_{i}X_{i}}{\delta(P_i-\bar{P}_i)}.
\end{align}
Picking $c_7=\frac{1}{4}$ and inverting the kinetic term, we obtain the propagator:
\begin{align}
   -{\hbar} \, \delta(P_i-\bar{P}_i) \longrightarrow \vcenter{\hbox{\includegraphics[height=1cm]{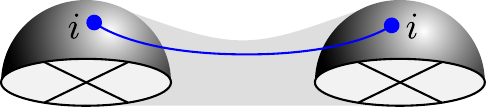}}}.
\end{align}
which has the topology of one-punctured $\mathbb{RP}^2$ times an interval. 

Notice that we have used $K_\id=1$ because the vacuum is invariant under parity. The sum over $i$ reduces to parity-even scalars.

One of the simplest new observables one can study now is the $\mathbb{RP}^2$ partition function. We can look for manifolds with a single $\mathbb{RP}^2$ boundary. However, there is no such smooth 3d manifold. Since the tensor model diagrams only produce smooth manifolds, this means that we get zero. If we compute the square of the $\mathbb{RP}^2$ partition function, however, we do get a non-zero answer. The wormhole with topology $\mathbb{RP}^2$ times an interval, for example, is one such contribution, even though the manifold is off-shell. This is consistent with the fact that the ensemble has a vanishing expectation value for this observable because $X_\id$ is equally likely to have a positive sign or a negative sign. This is to be contrasted with other models for computing the $\mathbb{RP}^2$ partition function in holography, where either a codimension-two \cite{Maloney:2016gsg} or codimension-one defect \cite{Wei:2024zez} is introduced to obtain a non-zero answer. In those proposals, the boundary is a single (non-orientable) CFT, and the bulk model is sensitive to details of the UV. (See also \cite{Verlinde:2015qfa,Nakayama:2016xvw,Lewkowycz:2016ukf} for related discussion on non-orientability in holography.)

\section*{Acknowledgements}

It is a pleasure to thank Scott Collier, Lorenz Eberhardt, Tom Hartman, Yikun Jiang, Jiaxin Qiao, Julian Sonner, Douglas Stanford, Zhencheng Wang, Zixia Wei, Gabriel Wong, and Mengyang Zhang for discussions. DLJ and LR acknowledge support by the Simons Investigator in Physics Award MP-SIP-0001737, U.S. Department of Energy grant DE-SC0007870, and Harvard University. DW acknowledges support by NSF grant PHY-2207659 and the Simons Collaboration on Celestial Holography. DLJ further thanks the Institute for Advanced Study for hospitality during the initial period of this work, and the IBM Einstein Fellow Fund for support.

\appendix

\section{Consistency checks}

In this appendix, we demonstrate with examples the consistency of the open-closed Virasoro TQFT partition functions and the surgery rules. 

\subsection*{Check 1}
The simplest example is the following:
\begin{align}\label{eq:V3xV3}
    \vcenter{\hbox{\includegraphics[height=3cm]{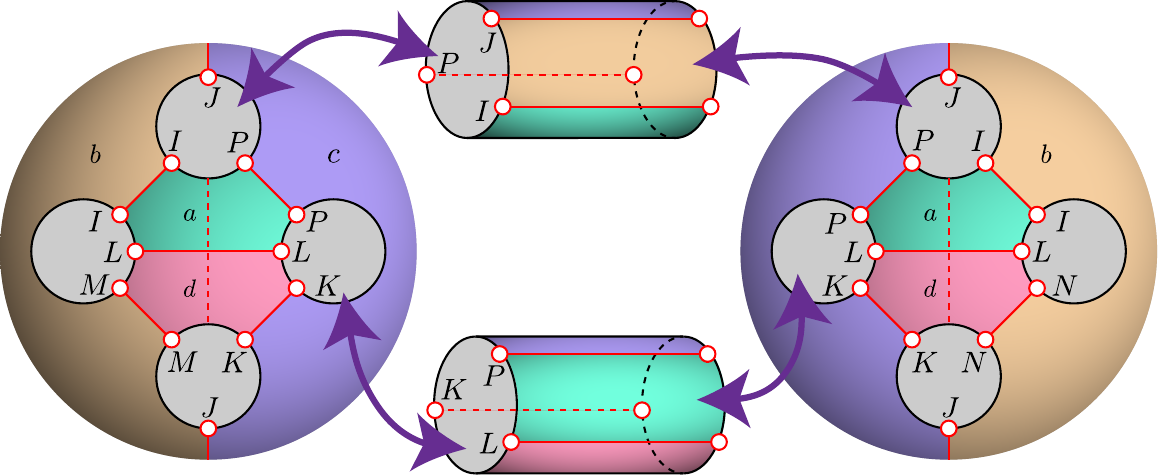}}}
    =\vcenter{\hbox{\includegraphics[height=2.5cm]{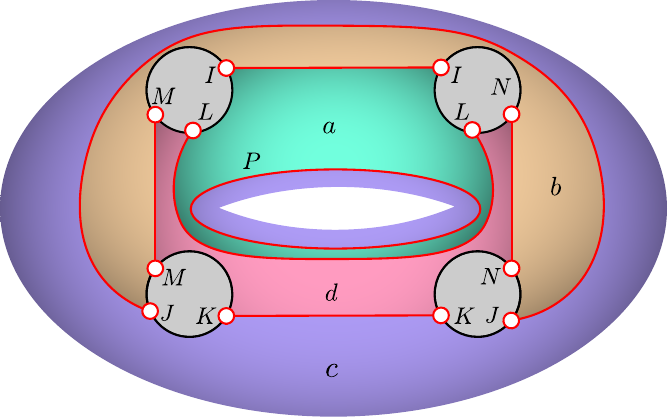}}}
    =\vcenter{\hbox{\includegraphics[height=2.5cm]{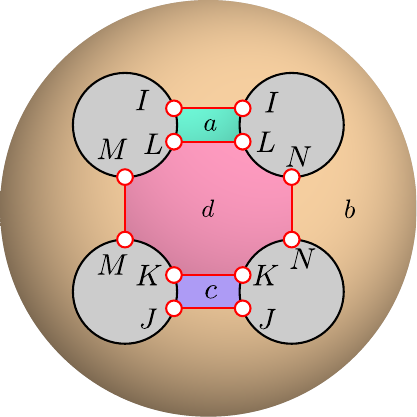}}}~.
\end{align}
First, we glue two open $6j$ manifolds \eqref{eq:V3cross} together using the $B$-propagator, which gives us a 3d solid torus, where a boundary Wilson loop wraps around the non-contractible boundary circle. In terms of formulae,
\begin{align}
    & C_0(IJP)C_0(KLP) 
    \times \frac{ \mathbb{F}_{M P} \begin{bmatrix} K & L \\ J & I \end{bmatrix} \mathbb{F}_{P N} \begin{bmatrix} K & J \\ L & I \end{bmatrix}}{g_ag_bg_cg_d\rho_0(P)\rho_0(N)C_0(IJP)C_0(KLP)C_0(ILN)C_0(JKN)} 
    \nn
    =\,&\frac{ \mathbb{F}_{M P} \begin{bmatrix} K & L \\ J & I \end{bmatrix} \mathbb{F}_{P N} \begin{bmatrix} K & J \\ L & I \end{bmatrix}}{g_ag_bg_cg_d\rho_0(P)\rho_0(N)C_0(ILN)C_0(JKN)}.
\end{align}
The annular surgery discussed in Section~\ref{ssec:sur} then removes this Wilson line along with the genus hole, creating a solid 3-ball. With the genus hole gone, the Wilson lines can be deformed to look like the open pillow manifold, as shown in the rightmost diagram above. Performing the integral corresponding to the surgery explicitly:
\begin{align}
    &\int \d P_P \,g_ag_c \rho_0(P)  
    \times \frac{ \mathbb{F}_{M P} \begin{bmatrix} K & L \\ J & I \end{bmatrix} \mathbb{F}_{P N} \begin{bmatrix} K & J \\ L & I \end{bmatrix}}{g_ag_bg_cg_d\rho_0(P)\rho_0(N)C_0(ILN)C_0(JKN)} 
    \nn
    =\,& \frac{1}{g_bg_d \rho_0(N) C_0(ILN) C_0(JKN) } \int \d P_P \,  \mathbb{F}_{MP} \begin{bmatrix} K & L \\ J & I \end{bmatrix} \mathbb{F}_{P N} \begin{bmatrix} K & J \\ L & I \end{bmatrix}  \nn
    =\,&  \frac{\delta(P_M-P_N)}{g_bg_d\rho_0(N) C_0(ILN) C_0(JKN)} ,
\end{align}
which is indeed the expression derived in \eqref{eq:V3diag}.

\subsection*{Check 2}

As another example, take two diagrams \eqref{eq:V4cross} and glue them with a $B$-propagator and a $D$-propagator as follows:
\begin{align}\label{eq:V4xV4xBD}
    \vcenter{\hbox{\includegraphics[height=3.5cm]{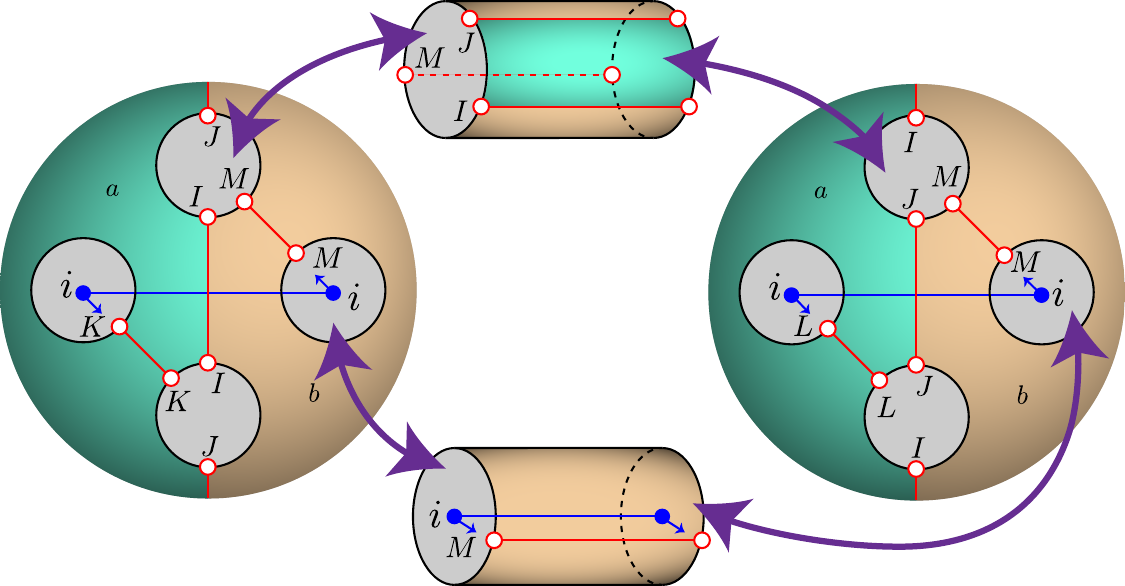}}}
    =\vcenter{\hbox{\includegraphics[height=2.5cm]{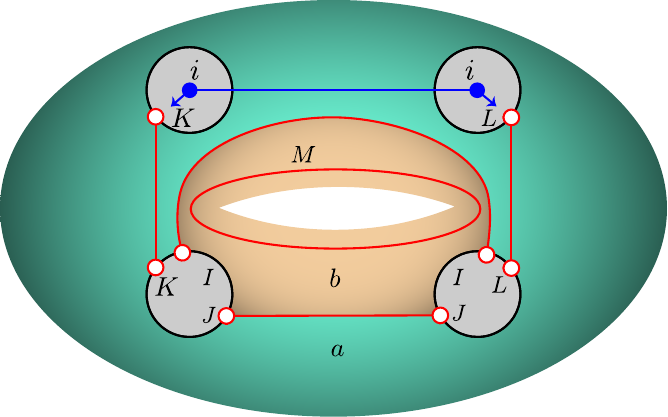}}}
    =\vcenter{\hbox{\includegraphics[height=2.5cm]{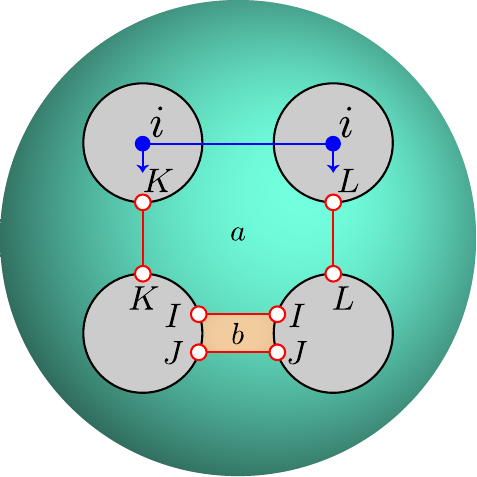}}}~.
\end{align}
The annular surgery again removes the boundary Wilson loop labeled by $M$ along with the genus hole. The framing has been included in this example (only shown at endpoints), which can be verified to work out correctly. In terms of formulae,  
\begin{align}
     &\int_0^{\infty}  \d P_{M} g_b^2\rho_0(M)     \times C_0(IJM)  C_0(i \ib M)\nn
     &\times\frac{(g_ag_b)^{-1}}{{\rho_0(K) C_0(i\ib K) C_0(IJK)}} \int_0^\infty \d P\, 
    \mathbb{F}_{M P}
    \begin{bmatrix}
        I & i \\
        J & \ib
    \end{bmatrix}
    \mathbb{F}_{P K}
    \begin{bmatrix}
        i & \ib \\
        I & J
    \end{bmatrix}
    \e^{-\i\pi(2h+h_M-h_i-\bar{h}_i-h_I-h_J)} \nn
     &\times\frac{(g_ag_b)^{-1}}{{\rho_0(L) C_0(\ib i L) C_0(IJL)}} \int_0^\infty \d P'\, 
    \mathbb{F}_{M P'}
    \begin{bmatrix}
        J & \ib \\
        I & i
    \end{bmatrix}
    \mathbb{F}_{P' L}
    \begin{bmatrix}
        \ib & i \\
        J & I
    \end{bmatrix}
    \e^{\i\pi(2h'+h_M-h_i-\bar{h}_i-h_I-h_J)} 
    \nn
    =\,& \frac{g_a^{-2}}{\rho_0(K)\rho_0(L) C_0(i\ib K) C_0(IJK) C_0(\ib i L) C_0(IJL)}  \nn
    & \times \int_0^\infty \d P\, \int_0^\infty \d P'\,
    \mathbb{F}_{P K}
    \begin{bmatrix}
        i & \ib \\
        I & J
    \end{bmatrix}
    \mathbb{F}_{P' L}
    \begin{bmatrix}
        \ib & i \\
        J & I
    \end{bmatrix}
    \e^{\i\pi(2h'-2h)} \nn
    & \times \rho_0(P') C_0(i I P') C_0(\ib J P') \int_0^{\infty}  \d P_{M} \mathbb{F}_{M P}
    \begin{bmatrix}
        I & i \\
        J & \ib
    \end{bmatrix} \mathbb{F}_{M P'}
    \begin{bmatrix}
        J & I \\
        \ib & i
    \end{bmatrix} 
    \nn
    =\,&\frac{g_a^{-2}}{\rho_0(K)\rho_0(L) C_0(i\ib K) C_0(IJK) C_0(\ib i L) C_0(IJL)}  \nn
    & \times \int_0^\infty \d P\, \int_0^\infty \d P'\,
    \mathbb{F}_{P K}
    \begin{bmatrix}
        i & \ib \\
        I & J
    \end{bmatrix}
    \mathbb{F}_{P' L}
    \begin{bmatrix}
        \ib & i \\
        J & I
    \end{bmatrix}
    \e^{\i\pi(2h'-2h)} \rho_0(P') C_0(i I P') C_0(\ib J P') \delta(P-P') 
    \nn
    =\,& \frac{g_a^{-2}}{\rho_0(K)\rho_0(L) C_0(i\ib K) C_0(IJK) C_0(\ib i L) C_0(IJL)} \nn
    &\times \int_0^\infty \d P\, \rho_0(P) C_0(i I P) C_0(\ib J P)
    \mathbb{F}_{P K}
    \begin{bmatrix}
        i & \ib \\
        I & J
    \end{bmatrix}
    \mathbb{F}_{P L}
    \begin{bmatrix}
        \ib & i \\
        J & I
    \end{bmatrix} 
    \nn
    =\,& \frac{\delta(P_K-P_L)}{g_a^2\rho_0(K)C_0(i\ib K) C_0(IJK)}. 
\end{align}

\subsection*{Check 3}

As a third example, we check explicitly that when two diagrams \eqref{eq:V5cross} are glued with a $B$-propagator and two $D$-propagators, we get \eqref{eq:V5diag1}:
\begin{align}
    & \int \d P_{J} g_a^2\rho_0(P_{J})  \int \d P_{K} g_a^2\rho_0(P_{K}) C_0(IJK)  C_0(j \jb J) C_0(k \bar k K)\nn
    &\times  \frac{\mathbb{K}_5[P_I,P_J;P_k,\bar{P}_k; P_K,P_i,P_j]}{g_a^{2} \rho_0(k)\rho_0(\bar{k})C_0(ijk)C_0(\ib\jb\bar{k})C_0(k\bar{k}K)}   \frac{\mathbb{K}_5[P_K,P_J;P_l,\bar{P}_l; P_I,P_k,P_j]}{g_a^{2} \rho_0(l)\rho_0(\bar{l})C_0(ljk)C_0(\bar l \jb\bar{k})C_0(l\bar{l}I)} \nn
    =\,&  \int \d P \d P' \frac{\mathbb{F}_{I k} \begin{bmatrix}
        \ib & P \\ i & j 
    \end{bmatrix} 
    \mathbb{F}_{P'l} \begin{bmatrix}
        k & j \\ \bar l & I
    \end{bmatrix} \e^{\frac{\i \pi}{2}(-h_i+\bar h_i -2h-\bar{h}_l+h_l+2h')}}{\rho_0(k)\rho_0(\bar k)C_0(ijk)C_0(\ib \jb \bar k) \rho_0(l) \rho_0(\bar l) C_0(kjl) C_0(\bar k \jb \bar l) C_0(l \bar l I)}  \nn
    &\times  \int \d P_J \d P_K\rho_0(P_J)   \rho_0(P_K) C_0(IJK) C_0(j \jb J) 
    \mathbb{F}_{JP} \begin{bmatrix}
        K & \jb \\ I & j
    \end{bmatrix} \mathbb{F}_{P \bar k} \begin{bmatrix}
        \ib & \jb \\ k & K
    \end{bmatrix} \mathbb{F}_{JP'} \begin{bmatrix}
        I & j \\ K & \jb
    \end{bmatrix} \mathbb{F}_{K \bar l} \begin{bmatrix}
        k & P' \\ \bar k & \jb
    \end{bmatrix} \nn
    =\,& \int \d P\d P' \frac{\mathbb{F}_{I k} \begin{bmatrix}
        \ib & P \\ i & j 
    \end{bmatrix} 
    \mathbb{F}_{P'l} \begin{bmatrix}
        k & j \\ \bar l & I
    \end{bmatrix} \e^{\frac{\i \pi}{2}(-h_i+\bar h_i -2h-\bar{h}_l+h_l+2h')}}{\rho_0(k)\rho_0(\bar k)C_0(ijk)C_0(\ib \jb \bar k) \rho_0(l) \rho_0(\bar l) C_0(kjl) C_0(\bar k \jb \bar l) C_0(l \bar l I)}  \nn
    &\times \int \d P_{K}  \rho_0(P_K) \rho_0(P) C_0(K \jb P) C_0(I j P) 
    \mathbb{F}_{P \bar k} \begin{bmatrix}
        \ib & \jb \\ k & K
    \end{bmatrix} \mathbb{F}_{K \bar l} \begin{bmatrix}
        k & P' \\ \bar k & \jb
    \end{bmatrix} \delta(P-P') \nn
    =\,&   \frac{ \delta(\bar{P}_l - \bar{P}_i)\e^{\frac{\i \pi}{2}(-h_i+h_l)}}{\rho_0(k) C_0(ijk) C_0(\ib \jb \bar k) \rho_0(l) \rho_0(\ib) C_0(kjl) C_0(l \ib I)} \int \d P \rho_0(P) C_0(I j P) \mathbb{F}_{I k} \begin{bmatrix}
        \ib & P \\ i & j 
    \end{bmatrix} 
    \mathbb{F}_{Pl} \begin{bmatrix}
        k & j \\ \ib & I
    \end{bmatrix} \nn
    =\,&   \frac{ \delta^2( P_l - P_i) }{\rho_0(i) \rho_0(\ib)  C_0(ijk) C_0(\ib \jb \bar k)  C_0(i \ib I)}.
\end{align}

\bibliographystyle{JHEP}
\bibliography{library}

\providecommand{\href}[2]{#2}\begingroup\raggedright\begin{thebibliography}{10}

\bibitem{Achucarro:1986uwr}
A.~Achucarro and P.~K. Townsend, \emph{{A Chern-Simons Action for Three-Dimensional anti-De Sitter Supergravity Theories}}, \href{https://doi.org/10.1016/0370-2693(86)90140-1}{\emph{Phys. Lett. B} {\bfseries 180} (1986) 89}.

\bibitem{Witten:1988hc}
E.~Witten, \emph{{(2+1)-Dimensional Gravity as an Exactly Soluble System}}, \href{https://doi.org/10.1016/0550-3213(88)90143-5}{\emph{Nucl. Phys. B} {\bfseries 311} (1988) 46}.

\bibitem{Collier:2023fwi}
S.~Collier, L.~Eberhardt and M.~Zhang, \emph{{Solving 3d gravity with Virasoro TQFT}}, \href{https://doi.org/10.21468/SciPostPhys.15.4.151}{\emph{SciPost Phys.} {\bfseries 15} (2023) 151} [\href{https://arxiv.org/abs/2304.13650}{{\ttfamily 2304.13650}}].

\bibitem{Collier:2024mgv}
S.~Collier, L.~Eberhardt and M.~Zhang, \emph{{3d gravity from Virasoro TQFT: Holography, wormholes and knots}}, \href{https://doi.org/10.21468/SciPostPhys.17.5.134}{\emph{SciPost Phys.} {\bfseries 17} (2024) 134} [\href{https://arxiv.org/abs/2401.13900}{{\ttfamily 2401.13900}}].

\bibitem{Witten:1988hf}
E.~Witten, \emph{{Quantum Field Theory and the Jones Polynomial}}, \href{https://doi.org/10.1007/BF01217730}{\emph{Commun. Math. Phys.} {\bfseries 121} (1989) 351}.

\bibitem{Verlinde:1989ua}
H.~L. Verlinde, \emph{{Conformal Field Theory, 2-$D$ Quantum Gravity and Quantization of Teichmuller Space}}, \href{https://doi.org/10.1016/0550-3213(90)90510-K}{\emph{Nucl. Phys. B} {\bfseries 337} (1990) 652}.

\bibitem{Maloney:2007ud}
A.~Maloney and E.~Witten, \emph{{Quantum Gravity Partition Functions in Three Dimensions}}, \href{https://doi.org/10.1007/JHEP02(2010)029}{\emph{JHEP} {\bfseries 02} (2010) 029} [\href{https://arxiv.org/abs/0712.0155}{{\ttfamily 0712.0155}}].

\bibitem{Keller:2014xba}
C.~A. Keller and A.~Maloney, \emph{{Poincare Series, 3D Gravity and CFT Spectroscopy}}, \href{https://doi.org/10.1007/JHEP02(2015)080}{\emph{JHEP} {\bfseries 02} (2015) 080} [\href{https://arxiv.org/abs/1407.6008}{{\ttfamily 1407.6008}}].

\bibitem{Cotler:2020ugk}
J.~Cotler and K.~Jensen, \emph{{AdS$_{3}$ gravity and random CFT}}, \href{https://doi.org/10.1007/JHEP04(2021)033}{\emph{JHEP} {\bfseries 04} (2021) 033} [\href{https://arxiv.org/abs/2006.08648}{{\ttfamily 2006.08648}}].

\bibitem{Cotler:2020hgz}
J.~Cotler and K.~Jensen, \emph{{AdS$_3$ wormholes from a modular bootstrap}}, \href{https://doi.org/10.1007/JHEP11(2020)058}{\emph{JHEP} {\bfseries 11} (2020) 058} [\href{https://arxiv.org/abs/2007.15653}{{\ttfamily 2007.15653}}].

\bibitem{DiUbaldo:2023qli}
G.~Di~Ubaldo and E.~Perlmutter, \emph{{AdS$_{3}$/RMT$_{2}$ duality}}, \href{https://doi.org/10.1007/JHEP12(2023)179}{\emph{JHEP} {\bfseries 12} (2023) 179} [\href{https://arxiv.org/abs/2307.03707}{{\ttfamily 2307.03707}}].

\bibitem{Boruch:2025ilr}
J.~Boruch, G.~Di~Ubaldo, F.~M. Haehl, E.~Perlmutter and M.~Rozali, \emph{{Modular-invariant random matrix theory and AdS${}_3$ wormholes}},  \href{https://arxiv.org/abs/2503.00101}{{\ttfamily 2503.00101}}.

\bibitem{deBoer:2025rct}
J.~de~Boer, J.~Kames-King and B.~Post, \emph{{Surgery and statistics in 3d gravity}},  \href{https://arxiv.org/abs/2506.04151}{{\ttfamily 2506.04151}}.

\bibitem{Belin:2020hea}
A.~Belin and J.~de~Boer, \emph{{Random statistics of OPE coefficients and Euclidean wormholes}}, \href{https://doi.org/10.1088/1361-6382/ac1082}{\emph{Class. Quant. Grav.} {\bfseries 38} (2021) 164001} [\href{https://arxiv.org/abs/2006.05499}{{\ttfamily 2006.05499}}].

\bibitem{Anous:2021caj}
T.~Anous, A.~Belin, J.~de~Boer and D.~Liska, \emph{{OPE statistics from higher-point crossing}}, \href{https://doi.org/10.1007/JHEP06(2022)102}{\emph{JHEP} {\bfseries 06} (2022) 102} [\href{https://arxiv.org/abs/2112.09143}{{\ttfamily 2112.09143}}].

\bibitem{Chandra:2022bqq}
J.~Chandra, S.~Collier, T.~Hartman and A.~Maloney, \emph{{Semiclassical 3D gravity as an average of large-c CFTs}}, \href{https://doi.org/10.1007/JHEP12(2022)069}{\emph{JHEP} {\bfseries 12} (2022) 069} [\href{https://arxiv.org/abs/2203.06511}{{\ttfamily 2203.06511}}].

\bibitem{deBoer:2023vsm}
J.~de~Boer, D.~Liska, B.~Post and M.~Sasieta, \emph{{A principle of maximum ignorance for semiclassical gravity}}, \href{https://doi.org/10.1007/JHEP02(2024)003}{\emph{JHEP} {\bfseries 2024} (2024) 003} [\href{https://arxiv.org/abs/2311.08132}{{\ttfamily 2311.08132}}].

\bibitem{deBoer:2024mqg}
J.~de~Boer, D.~Liska and B.~Post, \emph{{Multiboundary wormholes and OPE statistics}}, \href{https://doi.org/10.1007/JHEP10(2024)207}{\emph{JHEP} {\bfseries 10} (2024) 207} [\href{https://arxiv.org/abs/2405.13111}{{\ttfamily 2405.13111}}].

\bibitem{Wang:2025bcx}
D.~Wang, Z.~Wang and Z.~Wei, \emph{{Wormholes with Ends of the World}},  \href{https://arxiv.org/abs/2504.12278}{{\ttfamily 2504.12278}}.

\bibitem{Kraus:2016nwo}
P.~Kraus and A.~Maloney, \emph{{A cardy formula for three-point coefficients or how the black hole got its spots}}, \href{https://doi.org/10.1007/JHEP05(2017)160}{\emph{JHEP} {\bfseries 05} (2017) 160} [\href{https://arxiv.org/abs/1608.03284}{{\ttfamily 1608.03284}}].

\bibitem{Cardy:2017qhl}
J.~Cardy, A.~Maloney and H.~Maxfield, \emph{{A new handle on three-point coefficients: OPE asymptotics from genus two modular invariance}}, \href{https://doi.org/10.1007/JHEP10(2017)136}{\emph{JHEP} {\bfseries 10} (2017) 136} [\href{https://arxiv.org/abs/1705.05855}{{\ttfamily 1705.05855}}].

\bibitem{Collier:2019weq}
S.~Collier, A.~Maloney, H.~Maxfield and I.~Tsiares, \emph{{Universal dynamics of heavy operators in CFT$_{2}$}}, \href{https://doi.org/10.1007/JHEP07(2020)074}{\emph{JHEP} {\bfseries 07} (2020) 074} [\href{https://arxiv.org/abs/1912.00222}{{\ttfamily 1912.00222}}].

\bibitem{Belin:2021ryy}
A.~Belin, J.~de~Boer and D.~Liska, \emph{{Non-Gaussianities in the statistical distribution of heavy OPE coefficients and wormholes}}, \href{https://doi.org/10.1007/JHEP06(2022)116}{\emph{JHEP} {\bfseries 06} (2022) 116} [\href{https://arxiv.org/abs/2110.14649}{{\ttfamily 2110.14649}}].

\bibitem{Kusuki:2021gpt}
Y.~Kusuki, \emph{{Analytic bootstrap in 2D boundary conformal field theory: towards braneworld holography}}, \href{https://doi.org/10.1007/JHEP03(2022)161}{\emph{JHEP} {\bfseries 03} (2022) 161} [\href{https://arxiv.org/abs/2112.10984}{{\ttfamily 2112.10984}}].

\bibitem{Numasawa:2022cni}
T.~Numasawa and I.~Tsiares, \emph{{Universal dynamics of heavy operators in boundary CFT$_{2}$}}, \href{https://doi.org/10.1007/JHEP08(2022)156}{\emph{JHEP} {\bfseries 08} (2022) 156} [\href{https://arxiv.org/abs/2202.01633}{{\ttfamily 2202.01633}}].

\bibitem{Belin:2023efa}
A.~Belin, J.~de~Boer, D.~L. Jafferis, P.~Nayak and J.~Sonner, \emph{{Approximate CFTs and random tensor models}}, \href{https://doi.org/10.1007/JHEP09(2024)163}{\emph{JHEP} {\bfseries 09} (2024) 163} [\href{https://arxiv.org/abs/2308.03829}{{\ttfamily 2308.03829}}].

\bibitem{Jafferis:2024jkb}
D.~L. Jafferis, L.~Rozenberg and G.~Wong, \emph{{3d gravity as a random ensemble}}, \href{https://doi.org/10.1007/JHEP02(2025)208}{\emph{JHEP} {\bfseries 02} (2025) 208} [\href{https://arxiv.org/abs/2407.02649}{{\ttfamily 2407.02649}}].

\bibitem{Saad:2019lba}
P.~Saad, S.~H. Shenker and D.~Stanford, \emph{{JT gravity as a matrix integral}},  \href{https://arxiv.org/abs/1903.11115}{{\ttfamily 1903.11115}}.

\bibitem{Turaev:1992hq}
V.~G. Turaev and O.~Y. Viro, \emph{{State sum invariants of 3-manifolds and quantum 6 j-symbols }}, \href{https://doi.org/10.1016/0040-9383(92)90015-A}{\emph{Topology} {\bfseries 31} (1992) 865}.

\bibitem{Freidel:2005qe}
L.~Freidel, \emph{{Group field theory: An Overview}}, \href{https://doi.org/10.1007/s10773-005-8894-1}{\emph{Int. J. Theor. Phys.} {\bfseries 44} (2005) 1769} [\href{https://arxiv.org/abs/hep-th/0505016}{{\ttfamily hep-th/0505016}}].

\bibitem{Gurau:2011xp}
R.~Gurau and J.~P. Ryan, \emph{{Colored Tensor Models - a review}}, \href{https://doi.org/10.3842/SIGMA.2012.020}{\emph{SIGMA} {\bfseries 8} (2012) 020} [\href{https://arxiv.org/abs/1109.4812}{{\ttfamily 1109.4812}}].

\bibitem{Rivasseau2016}
V.~Rivasseau, \emph{Random tensors and quantum gravity}, \href{https://doi.org/10.3842/sigma.2016.069}{\emph{Symmetry, Integrability and Geometry: Methods and Applications} (2016) }.

\bibitem{Eberhardt:2023mrq}
L.~Eberhardt, \emph{{Notes on crossing transformations of Virasoro conformal blocks}},  \href{https://arxiv.org/abs/2309.11540}{{\ttfamily 2309.11540}}.

\bibitem{Lazaroiu:2000rk}
C.~I. Lazaroiu, \emph{{On the structure of open - closed topological field theory in two-dimensions}}, \href{https://doi.org/10.1016/S0550-3213(01)00135-3}{\emph{Nucl. Phys. B} {\bfseries 603} (2001) 497} [\href{https://arxiv.org/abs/hep-th/0010269}{{\ttfamily hep-th/0010269}}].

\bibitem{Lauda:2005wn}
A.~D. Lauda and H.~Pfeiffer, \emph{{Open-closed strings: Two-dimensional extended TQFTs and Frobenius algebras}},  \href{https://arxiv.org/abs/math/0510664}{{\ttfamily math/0510664}}.

\bibitem{Moore:2006dw}
G.~W. Moore and G.~Segal, \emph{{D-branes and K-theory in 2D topological field theory}},  \href{https://arxiv.org/abs/hep-th/0609042}{{\ttfamily hep-th/0609042}}.

\bibitem{Hung:2019bnq}
L.~Y. Hung and G.~Wong, \emph{{Entanglement branes and factorization in conformal field theory}}, \href{https://doi.org/10.1103/PhysRevD.104.026012}{\emph{Phys. Rev. D} {\bfseries 104} (2021) 026012} [\href{https://arxiv.org/abs/1912.11201}{{\ttfamily 1912.11201}}].

\bibitem{Brehm:2021wev}
E.~M. Brehm and I.~Runkel, \emph{{Lattice models from CFT on surfaces with holes: I. Torus partition function via two lattice cells}}, \href{https://doi.org/10.1088/1751-8121/ac6a91}{\emph{J. Phys. A} {\bfseries 55} (2022) 235001} [\href{https://arxiv.org/abs/2112.01563}{{\ttfamily 2112.01563}}].

\bibitem{Chen:2022wvy}
L.~Chen, K.~Ji, H.~Zhang, C.~Shen, R.~Wang, X.~Zeng et~al., \emph{{CFTD from TQFTD+1 via Holographic Tensor Network, and Precision Discretization of CFT2 }}, \href{https://doi.org/10.1103/PhysRevX.14.041033}{\emph{Phys. Rev. X} {\bfseries 14} (2024) 041033} [\href{https://arxiv.org/abs/2210.12127}{{\ttfamily 2210.12127}}].

\bibitem{Cheng:2023kxh}
G.~Cheng, L.~Chen, Z.-C. Gu and L.-Y. Hung, \emph{{Precision Reconstruction of Rational Conformal Field Theory from Exact Fixed-Point Tensor Network}}, \href{https://doi.org/10.1103/PhysRevX.15.011073}{\emph{Phys. Rev. X} {\bfseries 15} (2025) 011073} [\href{https://arxiv.org/abs/2311.18005}{{\ttfamily 2311.18005}}].

\bibitem{Chen:2024unp}
L.~Chen, L.-Y. Hung, Y.~Jiang and B.-X. Lao, \emph{{Deriving the non-perturbative gravitational dual of quantum Liouville theory from BCFT operator algebra}},  \href{https://arxiv.org/abs/2403.03179}{{\ttfamily 2403.03179}}.

\bibitem{Brehm:2024zun}
E.~M. Brehm and I.~Runkel, \emph{{Lattice models from CFT on surfaces with holes II: Cloaking boundary conditions and loop models}},  \href{https://arxiv.org/abs/2410.19938}{{\ttfamily 2410.19938}}.

\bibitem{Hung:2025vgs}
L.-Y. Hung, Y.~Jiang and B.-X. Lao, \emph{{Universal Structures and Emergent Geometry from Large-$c$ BCFT Ensemble}},  \href{https://arxiv.org/abs/2504.21660}{{\ttfamily 2504.21660}}.

\bibitem{Kravchuk:2021kwe}
P.~Kravchuk, J.~Qiao and S.~Rychkov, \emph{{Distributions in CFT. Part II. Minkowski space}}, \href{https://doi.org/10.1007/JHEP08(2021)094}{\emph{JHEP} {\bfseries 08} (2021) 094} [\href{https://arxiv.org/abs/2104.02090}{{\ttfamily 2104.02090}}].

\bibitem{Moore:1988uz}
G.~W. Moore and N.~Seiberg, \emph{{Polynomial Equations for Rational Conformal Field Theories}}, \href{https://doi.org/10.1016/0370-2693(88)91796-0}{\emph{Phys. Lett. B} {\bfseries 212} (1988) 451}.

\bibitem{Moore:1988qv}
G.~W. Moore and N.~Seiberg, \emph{{Classical and Quantum Conformal Field Theory}}, \href{https://doi.org/10.1007/BF01238857}{\emph{Commun. Math. Phys.} {\bfseries 123} (1989) 177}.

\bibitem{Cardy:1991tv}
J.~L. Cardy and D.~C. Lewellen, \emph{{Bulk and boundary operators in conformal field theory}}, \href{https://doi.org/10.1016/0370-2693(91)90828-E}{\emph{Phys. Lett. B} {\bfseries 259} (1991) 274}.

\bibitem{Lewellen:1991tb}
D.~C. Lewellen, \emph{{Sewing constraints for conformal field theories on surfaces with boundaries}}, \href{https://doi.org/10.1016/0550-3213(92)90370-Q}{\emph{Nucl. Phys. B} {\bfseries 372} (1992) 654}.

\bibitem{Ponsot:1999uf}
B.~Ponsot and J.~Teschner, \emph{{Liouville bootstrap via harmonic analysis on a noncompact quantum group}},  \href{https://arxiv.org/abs/hep-th/9911110}{{\ttfamily hep-th/9911110}}.

\bibitem{Ponsot:2000mt}
B.~Ponsot and J.~Teschner, \emph{{Clebsch-Gordan and Racah-Wigner coefficients for a continuous series of representations of U(q)(sl(2,R))}}, \href{https://doi.org/10.1007/PL00005590}{\emph{Commun. Math. Phys.} {\bfseries 224} (2001) 613} [\href{https://arxiv.org/abs/math/0007097}{{\ttfamily math/0007097}}].

\bibitem{Teschner:2012em}
J.~Teschner and G.~Vartanov, \emph{{6j symbols for the modular double, quantum hyperbolic geometry, and supersymmetric gauge theories}}, \href{https://doi.org/10.1007/s11005-014-0684-3}{\emph{Lett. Math. Phys.} {\bfseries 104} (2014) 527} [\href{https://arxiv.org/abs/1202.4698}{{\ttfamily 1202.4698}}].

\bibitem{Teschner:2013tqy}
J.~Teschner and G.~S. Vartanov, \emph{{Supersymmetric gauge theories, quantization of $\mathcal{M}_{\mathrm{flat}}$, and conformal field theory}}, \href{https://doi.org/10.4310/ATMP.2015.v19.n1.a1}{\emph{Adv. Theor. Math. Phys.} {\bfseries 19} (2015) 1} [\href{https://arxiv.org/abs/1302.3778}{{\ttfamily 1302.3778}}].

\bibitem{Fjelstad:2006aw}
J.~Fjelstad, J.~Fuchs, I.~Runkel and C.~Schweigert, \emph{{Uniqueness of open / closed rational CFT with given algebra of open states}}, \href{https://doi.org/10.4310/ATMP.2008.v12.n6.a4}{\emph{Adv. Theor. Math. Phys.} {\bfseries 12} (2008) 1283} [\href{https://arxiv.org/abs/hep-th/0612306}{{\ttfamily hep-th/0612306}}].

\bibitem{Dorn:1994xn}
H.~Dorn and H.~J. Otto, \emph{{Two and three point functions in Liouville theory}}, \href{https://doi.org/10.1016/0550-3213(94)00352-1}{\emph{Nucl. Phys. B} {\bfseries 429} (1994) 375} [\href{https://arxiv.org/abs/hep-th/9403141}{{\ttfamily hep-th/9403141}}].

\bibitem{Zamolodchikov:1995aa}
A.~B. Zamolodchikov and A.~B. Zamolodchikov, \emph{{Structure constants and conformal bootstrap in Liouville field theory}}, \href{https://doi.org/10.1016/0550-3213(96)00351-3}{\emph{Nucl. Phys. B} {\bfseries 477} (1996) 577} [\href{https://arxiv.org/abs/hep-th/9506136}{{\ttfamily hep-th/9506136}}].

\bibitem{Post:2024itb}
B.~Post and I.~Tsiares, \emph{{A non-rational Verlinde formula from Virasoro TQFT}}, \href{https://doi.org/10.1007/JHEP04(2025)015}{\emph{JHEP} {\bfseries 04} (2025) 015} [\href{https://arxiv.org/abs/2411.07285}{{\ttfamily 2411.07285}}].

\bibitem{Yan:2023rjh}
C.~Yan, \emph{{More on torus wormholes in 3d gravity}}, \href{https://doi.org/10.1007/JHEP11(2023)039}{\emph{JHEP} {\bfseries 11} (2023) 039} [\href{https://arxiv.org/abs/2305.10494}{{\ttfamily 2305.10494}}].

\bibitem{Takayanagi:2011zk}
T.~Takayanagi, \emph{{Holographic Dual of BCFT}}, \href{https://doi.org/10.1103/PhysRevLett.107.101602}{\emph{Phys. Rev. Lett.} {\bfseries 107} (2011) 101602} [\href{https://arxiv.org/abs/1105.5165}{{\ttfamily 1105.5165}}].

\bibitem{Felder:1999cv}
G.~Felder, J.~Frohlich, J.~Fuchs and C.~Schweigert, \emph{{Conformal boundary conditions and three-dimensional topological field theory}}, \href{https://doi.org/10.1103/PhysRevLett.84.1659}{\emph{Phys. Rev. Lett.} {\bfseries 84} (2000) 1659} [\href{https://arxiv.org/abs/hep-th/9909140}{{\ttfamily hep-th/9909140}}].

\bibitem{Felder:1999mq}
G.~Felder, J.~Frohlich, J.~Fuchs and C.~Schweigert, \emph{{Correlation functions and boundary conditions in RCFT and three-dimensional topology}}, \href{https://doi.org/10.1023/A:1014903315415}{\emph{Compos. Math.} {\bfseries 131} (2002) 189} [\href{https://arxiv.org/abs/hep-th/9912239}{{\ttfamily hep-th/9912239}}].

\bibitem{Fuchs:2002cm}
J.~Fuchs, I.~Runkel and C.~Schweigert, \emph{{TFT construction of RCFT correlators 1. Partition functions}}, \href{https://doi.org/10.1016/S0550-3213(02)00744-7}{\emph{Nucl. Phys. B} {\bfseries 646} (2002) 353} [\href{https://arxiv.org/abs/hep-th/0204148}{{\ttfamily hep-th/0204148}}].

\bibitem{Fuchs:2003id}
J.~Fuchs, I.~Runkel and C.~Schweigert, \emph{{TFT construction of RCFT correlators. 2. Unoriented world sheets}}, \href{https://doi.org/10.1016/j.nuclphysb.2003.11.026}{\emph{Nucl. Phys. B} {\bfseries 678} (2004) 511} [\href{https://arxiv.org/abs/hep-th/0306164}{{\ttfamily hep-th/0306164}}].

\bibitem{Fuchs:2004dz}
J.~Fuchs, I.~Runkel and C.~Schweigert, \emph{{TFT construction of RCFT correlators. 3. Simple currents}}, \href{https://doi.org/10.1016/j.nuclphysb.2004.05.014}{\emph{Nucl. Phys. B} {\bfseries 694} (2004) 277} [\href{https://arxiv.org/abs/hep-th/0403157}{{\ttfamily hep-th/0403157}}].

\bibitem{Fuchs:2004xi}
J.~Fuchs, I.~Runkel and C.~Schweigert, \emph{{TFT construction of RCFT correlators IV: Structure constants and correlation functions}}, \href{https://doi.org/10.1016/j.nuclphysb.2005.03.018}{\emph{Nucl. Phys. B} {\bfseries 715} (2005) 539} [\href{https://arxiv.org/abs/hep-th/0412290}{{\ttfamily hep-th/0412290}}].

\bibitem{Fjelstad:2005ua}
J.~Fjelstad, J.~Fuchs, I.~Runkel and C.~Schweigert, \emph{{TFT construction of RCFT correlators. V. Proof of modular invariance and factorisation}}, {\emph{Theor. Appl. Categor.} {\bfseries 16} (2006) 342} [\href{https://arxiv.org/abs/hep-th/0503194}{{\ttfamily hep-th/0503194}}].

\bibitem{Frohlich:2006ch}
J.~Frohlich, J.~Fuchs, I.~Runkel and C.~Schweigert, \emph{{Duality and defects in rational conformal field theory}}, \href{https://doi.org/10.1016/j.nuclphysb.2006.11.017}{\emph{Nucl. Phys. B} {\bfseries 763} (2007) 354} [\href{https://arxiv.org/abs/hep-th/0607247}{{\ttfamily hep-th/0607247}}].

\bibitem{Kong:2009inh}
L.~Kong and I.~Runkel, \emph{{Cardy algebras and sewing constraints. I.}}, \href{https://doi.org/10.1007/s00220-009-0901-6}{\emph{Commun. Math. Phys.} {\bfseries 292} (2009) 871} [\href{https://arxiv.org/abs/0807.3356}{{\ttfamily 0807.3356}}].

\bibitem{Kong:2013gca}
L.~Kong, Q.~Li and I.~Runkel, \emph{{Cardy algebras and sewing constraints, II}}, \href{https://doi.org/10.1016/j.aim.2014.05.020}{\emph{Adv. Math.} {\bfseries 262} (2014) 604} [\href{https://arxiv.org/abs/1310.1875}{{\ttfamily 1310.1875}}].

\bibitem{Traube:2020pfg}
M.~Traube, \emph{{Cardy Algebras, Sewing Constraints and String-Nets}}, \href{https://doi.org/10.1007/s00220-021-04286-6}{\emph{Commun. Math. Phys.} {\bfseries 390} (2022) 67} [\href{https://arxiv.org/abs/2009.11895}{{\ttfamily 2009.11895}}].

\bibitem{Hartman:2025cyj}
T.~Hartman, \emph{{Conformal Turaev-Viro Theory}},  \href{https://arxiv.org/abs/2507.11652}{{\ttfamily 2507.11652}}.

\bibitem{Hartman:2025ula}
T.~Hartman, \emph{{Triangulating quantum gravity in AdS$_3$}},  \href{https://arxiv.org/abs/2507.12696}{{\ttfamily 2507.12696}}.

\bibitem{Runkel:1998he}
I.~Runkel, \emph{{Boundary structure constants for the A series Virasoro minimal models}}, \href{https://doi.org/10.1016/S0550-3213(99)00125-X}{\emph{Nucl. Phys. B} {\bfseries 549} (1999) 563} [\href{https://arxiv.org/abs/hep-th/9811178}{{\ttfamily hep-th/9811178}}].

\bibitem{Fateev:2000ik}
V.~Fateev, A.~B. Zamolodchikov and A.~B. Zamolodchikov, \emph{{Boundary Liouville field theory. 1. Boundary state and boundary two point function}},  \href{https://arxiv.org/abs/hep-th/0001012}{{\ttfamily hep-th/0001012}}.

\bibitem{Teschner:2000md}
J.~Teschner, \emph{{Remarks on Liouville theory with boundary}}, \href{https://doi.org/10.22323/1.006.0041}{\emph{PoS} {\bfseries tmr2000} (2000) 041} [\href{https://arxiv.org/abs/hep-th/0009138}{{\ttfamily hep-th/0009138}}].

\bibitem{Zamolodchikov:2001ah}
A.~B. Zamolodchikov and A.~B. Zamolodchikov, \emph{{Liouville field theory on a pseudosphere}},  \href{https://arxiv.org/abs/hep-th/0101152}{{\ttfamily hep-th/0101152}}.

\bibitem{Collier:2023cyw}
S.~Collier, L.~Eberhardt, B.~M{\"u}hlmann and V.~A. Rodriguez, \emph{{The Virasoro minimal string}}, \href{https://doi.org/10.21468/SciPostPhys.16.2.057}{\emph{SciPost Phys.} {\bfseries 16} (2024) 057} [\href{https://arxiv.org/abs/2309.10846}{{\ttfamily 2309.10846}}].

\bibitem{Maloney:2015ina}
A.~Maloney, \emph{{Geometric Microstates for the Three Dimensional Black Hole?}},  \href{https://arxiv.org/abs/1508.04079}{{\ttfamily 1508.04079}}.

\bibitem{Eberhardt:2022wlc}
L.~Eberhardt, \emph{{Off-shell Partition Functions in 3d Gravity}}, \href{https://doi.org/10.1007/s00220-024-04963-2}{\emph{Commun. Math. Phys.} {\bfseries 405} (2024) 76} [\href{https://arxiv.org/abs/2204.09789}{{\ttfamily 2204.09789}}].

\bibitem{Fujita:2011fp}
M.~Fujita, T.~Takayanagi and E.~Tonni, \emph{{Aspects of AdS/BCFT}}, \href{https://doi.org/10.1007/JHEP11(2011)043}{\emph{JHEP} {\bfseries 11} (2011) 043} [\href{https://arxiv.org/abs/1108.5152}{{\ttfamily 1108.5152}}].

\bibitem{Geng22}
H.~Geng, \emph{{Aspects of AdS$_{2}$ quantum gravity and the Karch-Randall braneworld}}, \href{https://doi.org/10.1007/JHEP09(2022)024}{\emph{JHEP} {\bfseries 09} (2022) 024} [\href{https://arxiv.org/abs/2206.11277}{{\ttfamily 2206.11277}}].

\bibitem{VanRaamsdonk:2018zws}
M.~Van~Raamsdonk, \emph{{Building up spacetime with quantum entanglement II: It from BC-bit}},  \href{https://arxiv.org/abs/1809.01197}{{\ttfamily 1809.01197}}.

\bibitem{Hung:2024gma}
L.-Y. Hung and Y.~Jiang, \emph{{Building up quantum spacetimes with BCFT Legos}},  \href{https://arxiv.org/abs/2404.00877}{{\ttfamily 2404.00877}}.

\bibitem{Bao:2024ixc}
N.~Bao, L.-Y. Hung, Y.~Jiang and Z.~Liu, \emph{{QG from SymQRG: AdS$_3$/CFT$_2$ Correspondence as Topological Symmetry-Preserving Quantum RG Flow}},  \href{https://arxiv.org/abs/2412.12045}{{\ttfamily 2412.12045}}.

\bibitem{Roberts1995}
J.~Roberts, \emph{{Skein theory and Turaev-Viro invariants}}, \href{https://doi.org/10.1016/0040-9383(94)00053-0}{\emph{Topology} {\bfseries 34} (1995) }.

\bibitem{Pradisi:1995pp}
G.~Pradisi, A.~Sagnotti and Y.~S. Stanev, \emph{{The Open descendants of nondiagonal SU(2) WZW models}}, \href{https://doi.org/10.1016/0370-2693(95)00840-H}{\emph{Phys. Lett. B} {\bfseries 356} (1995) 230} [\href{https://arxiv.org/abs/hep-th/9506014}{{\ttfamily hep-th/9506014}}].

\bibitem{Stanev:2001na}
Y.~S. Stanev, \emph{{Two-dimensional conformal field theory on open and unoriented surfaces}},  in \emph{{4th SIGRAV Graduate School on Contemporary Relativity and Gravitational Physics and 2001 School on Algebraic Geometry and Physics: Geometry and Physics of Branes (SAGP 2001)}}, pp.~39--85, 12, 2001, \href{https://arxiv.org/abs/hep-th/0112222}{{\ttfamily hep-th/0112222}}.

\bibitem{Tsiares:2020ewp}
I.~Tsiares, \emph{{Universal Dynamics in Non-Orientable CFT$_2$}},  \href{https://arxiv.org/abs/2011.09250}{{\ttfamily 2011.09250}}.

\bibitem{Fioravanti:1993hf}
D.~Fioravanti, G.~Pradisi and A.~Sagnotti, \emph{{Sewing constraints and nonorientable open strings}}, \href{https://doi.org/10.1016/0370-2693(94)90255-0}{\emph{Phys. Lett. B} {\bfseries 321} (1994) 349} [\href{https://arxiv.org/abs/hep-th/9311183}{{\ttfamily hep-th/9311183}}].

\bibitem{Maloney:2016gsg}
A.~Maloney and S.~F. Ross, \emph{{Holography on Non-Orientable Surfaces}}, \href{https://doi.org/10.1088/0264-9381/33/18/185006}{\emph{Class. Quant. Grav.} {\bfseries 33} (2016) 185006} [\href{https://arxiv.org/abs/1603.04426}{{\ttfamily 1603.04426}}].

\bibitem{Wei:2024zez}
Z.~Wei, \emph{{Holographic dual of crosscap conformal field theory}}, \href{https://doi.org/10.1007/JHEP03(2025)086}{\emph{JHEP} {\bfseries 03} (2025) 086} [\href{https://arxiv.org/abs/2405.03755}{{\ttfamily 2405.03755}}].

\bibitem{Verlinde:2015qfa}
H.~Verlinde, \emph{{Poking Holes in AdS/CFT: Bulk Fields from Boundary States}},  \href{https://arxiv.org/abs/1505.05069}{{\ttfamily 1505.05069}}.

\bibitem{Nakayama:2016xvw}
Y.~Nakayama and H.~Ooguri, \emph{{Bulk Local States and Crosscaps in Holographic CFT}}, \href{https://doi.org/10.1007/JHEP10(2016)085}{\emph{JHEP} {\bfseries 10} (2016) 085} [\href{https://arxiv.org/abs/1605.00334}{{\ttfamily 1605.00334}}].

\bibitem{Lewkowycz:2016ukf}
A.~Lewkowycz, G.~J. Turiaci and H.~Verlinde, \emph{{A CFT Perspective on Gravitational Dressing and Bulk Locality}}, \href{https://doi.org/10.1007/JHEP01(2017)004}{\emph{JHEP} {\bfseries 01} (2017) 004} [\href{https://arxiv.org/abs/1608.08977}{{\ttfamily 1608.08977}}].

\end{thebibliography}\endgroup

\end{document}